\documentclass[twocolumn,showkeys,amsmath,amssymb,pre]{revtex4}

\usepackage{color}
\usepackage{graphicx}
\usepackage{bbm}
\usepackage{amssymb,amsbsy,amsmath}
\newcommand{\rd}{{\mathrm d}}
\newcommand{\re}{{\mathrm e}}
\newcommand{\ri}{{\mathrm i}}
\newcommand{\kB}{k_{\rm B}}

\begin{document}

\title[Periodic thermodynamics]
      {Periodic thermodynamics of the Rabi model with circular polarization \\
       for arbitrary spin quantum numbers}

\author{Heinz-J\"urgen Schmidt$^1$,
	J\"urgen Schnack$^2$,
	and Martin Holthaus$^3$}
	
\affiliation{$^1$Universit\"at Osnabr\"uck, Fachbereich Physik,
 	D-49069 Osnabr\"uck, Germany}
\affiliation{$^2$Universit\"at Bielefeld, Fakult\"at f\"ur Physik,
	D-33501 Bielefeld, Germany}
\affiliation{$^3$Carl von Ossietzky Universit\"at, Institut f\"ur Physik,
	D-26111 Oldenburg, Germany}


\begin{abstract}
We consider a spin~$s$ subjected to both a static and an orthogonally applied
oscillating, circularly polarized magnetic field while being coupled to a
heat bath, and analytically determine the quasi\-stationary distribution of
its Floquet-state occupation probabilities for arbitrarily strong driving.
This distribution is shown to be Boltzmannian with a quasitemperature which
is different from the temperature of the bath, and independent of the
spin quantum number. We discover a remarkable formal analogy between the
quasithermal magnetism of the nonequilibrium steady state of a driven ideal
paramagnetic material, and the usual thermal paramagnetism. Nonetheless, the
response of such a material to the combined fields is predicted to show several
unexpected features, even allowing one to turn a paramagnet into a diamagnet
under strong driving. Thus, we argue that experimental measurements of this
response may provide key paradigms for the emerging field of periodic
thermodynamics.
\end{abstract}


\maketitle


\section{Introduction}
\label{sec:I}

A quantum system governed by an explicitly time-dependent Hamiltonian $H(t)$
which varies {\em periodically\/} with time~$t$, such that
\begin{equation}
	H(t) = H(t+T) \; ,
\end{equation}	
possesses a complete set of {\em Floquet states\/}, that is, of solutions to
the time-dependent Schr\"odinger equation having the particular form
\begin{equation}
	|\psi_n(t)\rangle = |u_n(t)\rangle \exp(-\ri\varepsilon_n t) \; .
\label{eq:FST}
\end{equation}
The {\em Floquet functions\/} $| u_n(t) \rangle$ share the $T$-periodic
time dependence of their Hamiltonian,
\begin{equation}
	| u_n(t) \rangle = | u_n(t+T) \rangle \; ;
\end{equation}	
the quantities $\varepsilon_n$, which accompany their time evolution in the
same manner as energy eigenvalues accompany the evolution of unperturbed
energy eigenstates, are known as {\em quasienergies\/}~\cite{Zeldovich66,
Sambe73,FainshteinEtAl78}. Here we assume that the quasienergies constitute a
pure point spectrum, associated with square-integrable Floquet states in
the system's Hilbert space~${\mathcal H}_S$; we also adopt a system of units
such that both the Planck constant~$\hbar$ and the Boltzmann constant $\kB$
are set to one.

Evidently the factorization of a Floquet state~(\ref{eq:FST}) into a Floquet
function and an exponential of a phase which grows linearly in time is not
unique: Defining $\omega = 2\pi/T$, and taking an arbitrary, positive or
negative integer~$\nu$, one has
\begin{equation}
	|u_n(t)\rangle \exp(-\ri\varepsilon_n t) =
	|u_n(t) \re^{\ri\nu\omega t}\rangle
	\exp\!\big(-\ri[ \varepsilon_n + \nu\omega ]t\big) \; ,
\end{equation}
where $|u_n(t) \re^{\ri\nu\omega t}\rangle$ again is a $T$-periodic Floquet
function, representing the same Floquet state as $|u_n(t)\rangle$. Therefore,
a quasienergy is not to be regarded as just a single number equipped with
the dimension of energy, but rather as an infinite set of equivalent
representatives,
\begin{equation}
	[\varepsilon_n ] \equiv \{ \varepsilon_n + \nu\omega \; | \;
	\nu \in {\mathbbm Z} \} \; ,
\label{eq:QLA}
\end{equation}
where the choice of the ``canonical representative'' distinguished by setting
$\nu = 0$ is a matter of convention.

The significance of these Floquet states~(\ref{eq:FST}) rests in the fact that,
as long as the Hamiltonian depends on time in a strictly $T$-periodic manner,
every solution $| \psi(t) \rangle$ to the time-dependent Schr\"odinger equation
can be expanded with respect to the Floquet basis,
\begin{equation}
	| \psi(t) \rangle = \sum_n c_n \,  	
	|u_n(t)\rangle \exp(-\ri\varepsilon_n t) \; ,
\end{equation}
where the coefficients $c_n$ do not depend on time. Hence, the Floquet states
propagate with constant occupation probabilities~$| c_n |^2$, despite the
presence of a time-periodic drive. Under conditions of perfectly coherent
time evolution these coefficients $c_n$ would be determined solely by the
system's state at the moment the periodic drive is turned on. However, if the
periodically driven system is interacting with an environment, as it happens
in many cases of experimental interest~\cite{BlumelEtAl91,GrifoniHanggi98,
GasparinettiEtAl13,StaceEtAl13,ZhangEtAl17,ChoiEtAl17}, that environment may
continuously induce transitions among the system's Floquet states, to the
effect that a quasi\-stationary distribution $\{ p_n \}$ of Floquet-state
occupation probabilities establishes itself which contains no memory of the
initial state, and the question emerges how to quantify this distribution.

In a short programmatic note entitled ``Periodic Thermodynamics'', Kohn
has drawn attention to such quasi\-stationary Floquet-state distributions
$\{ p_n \}$, emphasizing that they should be less universal than usual
distributions characterizing thermal equilibrium, depending on the very form
of the system's interaction with its environment~\cite{Kohn01}. In an earlier
pioneering study, Breuer {\em et al.\/} had already calculated these
distributions for time-periodically forced oscillators coupled to a thermal
oscillator bath~\cite{BreuerEtAl00}. For the particular case of a linearly
forced {\em harmonic\/} oscillator these authors have shown that the
Floquet-state distribution remains a Boltzmann distribution parametrized by the
temperature of the heat bath, whereas it becomes rather more complicated in
the case of forced {\em anharmonic\/} oscillators. These investigations have
been extended later by Ketzmerick and Wustmann, who have demonstrated that
structures found in the phase space of classical forced anharmonic oscillators
leave their distinct traces in the quasistationary Floquet-state distributions
of their quantum counterparts~\cite{KetzmerickWustmann10}. To date, a great
variety of different individual aspects of the ``periodic thermodynamics''
envisioned by Kohn has been discussed in the literature~\cite{HoneEtAl09,
BulnesCuetaraEtAl15,ShiraiEtAl15,Liu15,IadecolaEtAl15a,IadecolaEtAl15,
SeetharamEtAl15,VorbergEtAl15,VajnaEtAl16,RestrepoEtAl16,LazaridesMoessner17,
SeetharamEtAl19}, but a coherent overall picture is still lacking.

In this situation it seems advisable to resort to models which are sufficiently
simple to admit analytical solutions and thus to unravel salient features on
the one hand, and which actually open up meaningful perspectives for
groundbreaking novel laboratory experiments on the other. To this end, in the
present work we consider a spin~$s$ exposed to both a static magnetic field and
an oscillating, circularly polarized magnetic field applied perpendicularly to
the static one, as in the classic Rabi set-up~\cite{Rabi37}, and coupled to a
thermal bath of harmonic oscillators. The experimental measurement of the
thermal paramagnetism resulting from magnetic moments subjected to a static
field alone has a long and successful history~\cite{Brillouin27,Henry52},
having become a standard topic in textbooks on Statistical
Physics~\cite{FowlerGuggenheim39,Pathria11}. We argue that a future generation
of such experiments, including both a static and a strong oscillating field,
may set further milestones towards the development of full-fledged periodic
thermodynamics.

We proceed as follows: In Sec.~\ref{sec:II} we collect the necessary technical
tools, starting with a brief summary of the golden-rule approach to
time-periodically driven open quantum systems in the form developed by Breuer
{\em et al.\/}~\cite{BreuerEtAl00}, thereby establishing our notation.
We also sketch a technique which enables one to ``lift'' a solution to the
Schr\"odinger equation for a spin $s = 1/2$ in a time-varying magnetic field
to general~$s$. In Sec.~\ref{sec:III} we discuss the Floquet states for spins
in a circularly polarized driving field, obtaining the states for general~$s$
from those for $s = 1/2$ with the help of the lifting procedure. In
Sec.~\ref{sec:IV} we compute the quasistationary Floquet-state distribution
for driven spins under the assumption that the spectral density of the heat
bath be constant, and show that this distribution is Boltzmannian with a
quasitemperature which is {\em different\/} from the actual bath temperature;
the dependence of this quasitemperature on the system parameters is discussed
in some detail. In Sec.~\ref{sec:V} we determine the magnetization of a spin
system which is subjected to both a static and an orthogonally applied,
circularly polarized magnetic field while being coupled to a heat bath. To this
end, we first establish a general formula for the ensuing magnetization by
means of another systematic use of the lifting technique, and then show that
the resulting expression can be interpreted as a derivative of a partition
function based on both the quasi\-temperature and the system's quasienergies,
in perfect formal analogy to the textbook treatment of paramagnetism in the
absence of time-periodic driving; these insights are exploited for elucidating
the response of an ideal paramagnet to a circularly polarized driving field.
In Sec.~\ref{sec:VI} we consider the rate of energy dissipated by the driven
spins into the bath, thus generalizing results derived previously for $s = 1/2$
in Ref.~\cite{LangemeyerHolthaus14}. In Sec.~\ref{sec:VII} we summarize and
discuss our main findings, emphasizing the possible knowledge gain to be
derived from future measurements of paramagnetic response to strong
time-periodic forcing, carried out along the lines drawn in the present work.

\section{Technical Tools}
\label{sec:II}

\subsection{Golden-rule approach to open driven systems}
\label{sec:2a}

Let us consider a quantum system evolving according to a $T$-periodic
Hamiltonian~$H(t)$ on a Hilbert space ${\mathcal H}_S$ which is perturbed by a
time-independent operator~$V$. Then the transition matrix element connecting
an initial Floquet function $|u_i(t)\rangle$ to a final Floquet function
$|u_f(t)\rangle$ can be expanded into a Fourier series,
\begin{equation}
	\langle u_f(t) | \, V \, | u_i(t) \rangle
	= \sum_{\ell\in{\mathbbm Z}} \, V_{fi}^{(\ell)}
	\exp(\ri\ell\omega t) \; ,
\label{eq:FDV}	
\end{equation}
and consequently the ``golden rule'' for the rate of transitions $\Gamma_{fi}$
from a Floquet state labeled~$i$ to a Floquet state~$f$ is written
as~\cite{LangemeyerHolthaus14}
\begin{equation}
	\Gamma_{fi} = 2\pi \sum_{\ell\in{\mathbbm Z}} \,
	| V_{fi}^{(\ell)} |^2 \,
	\delta(\omega_{fi}^{(\ell)}) \; ,
\end{equation}
where
\begin{equation}
	\omega_{fi}^{(\ell)} = \varepsilon_f - \varepsilon_i + \ell\omega
	\; .
\label{eq:OFI}
\end{equation}	
Thus, a transition among Floquet states is not simply associated with only
one single frequency, but rather with a set of frequencies spaced by integer
multiples of the driving frequency~$\omega$, reflecting the ladder-like nature
of the system's quasienergies~(\ref{eq:QLA}); this is one of the sources of the
peculiarities which distinguish periodic thermodynamics from usual equilibrium
thermodynamics~\cite{Kohn01,BreuerEtAl00}.

Let us now assume that, instead of merely being perturbed by~$V$, the
periodically driven system is coupled to a heat bath, described by a
Hamiltonian $H_{\rm bath}$ acting on a Hilbert space ${\mathcal H}_{B}$,
so that the total Hamiltonian on the composite Hilbert space
${\mathcal H}_S \otimes {\mathcal H}_B$ takes the form
\begin{equation}
	H_{\rm total}(t) = H(t) \otimes {\mathbbm 1}
	+ {\mathbbm 1} \otimes H_{\rm bath} + H_{\rm int} \; .
\end{equation}
Stipulating further that the interaction Hamiltonian $H_{\rm int}$
factorizes according to
\begin{equation}
	H_{\rm int} = V \otimes W \; ,
\label{eq:HVW}
\end{equation}
the golden rule can be applied to joint transitions from Floquet states~$i$
to Floquet states~$f$ of the system accompanied by transitions from bath
eigenstates~$n$ with energy $E_n$ to other bath eigenstates~$m$ with energy
$E_m$, acquiring the form
\begin{equation}
	\Gamma_{fi}^{mn} = 2\pi \sum_{\ell\in{\mathbbm Z}} \,
	| V_{fi}^{(\ell)} |^2 \, | W_{mn} |^2 \,
	 \delta(E_m - E_n + \omega_{fi}^{(\ell)}) \; .
\label{eq:GOR}
\end{equation}
Moreover, following Breuer {\em et al.\/}~\cite{BreuerEtAl00}, let us
consider a bath consisting of thermally occupied harmonic oscillators,
and an interaction of the  prototypical form
\begin{equation}
	W = \sum_{\widetilde\omega} \left(
	b_{\widetilde\omega}^{\phantom\dagger} + b_{\widetilde\omega}^\dagger
	\right) \; ,	
\label{eq:PTF}
\end{equation}
where $b_{\widetilde\omega}^{\phantom\dagger}$ ($b_{\widetilde\omega}^\dagger$)
is the annihilation (creation) operator pertaining to a bath oscillator of
frequency~$\widetilde{\omega}$. One could also multiply~$W$ by a
function~$g(\widetilde{\omega})$ specifying a frequency-dependent coupling
strength, but this function could be absorbed in the spectral density of the
bath introduced later, and therefore will not be used here.

We now have to distinguish two cases: If $E_n - E_m = \widetilde{\omega} > 0$,
so that the bath is de-excited and transfers energy to the system, the required
annihilation-operator matrix element reads
\begin{equation}
	W_{mn} = \sqrt{n(\widetilde{\omega})} \; ,
\end{equation}
where $n(\widetilde{\omega})$ is the occupation number of a bath oscillator
with frequency~$\widetilde{\omega}$, and the square $| W_{mn} |^2$ entering
the golden rule~(\ref{eq:GOR}) has to be replaced by the thermal avarage
\begin{equation}
	N(\widetilde{\omega})
	\equiv \langle n(\widetilde{\omega}) \rangle
	= \frac{1}{\exp(\beta\widetilde{\omega}) - 1} \; ,
\label{eq:NOP}	
\end{equation}
with~$\beta$ denoting the inverse bath temperature. Conversely, if
$E_n - E_m = \widetilde{\omega} < 0$ so that the system loses energy to the
bath and a bath phonon is created, one has
\begin{equation}
	W_{mn} = \sqrt{n(-\widetilde{\omega}) + 1} \; ,
\end{equation}
giving
\begin{equation}
	N(\widetilde{\omega})
	\equiv \langle n(-\widetilde{\omega}) \rangle + 1
	= \frac{1}{1 - \exp(\beta\widetilde{\omega})} \; .
\label{eq:NOM}	
\end{equation}
Finally, let $J(\widetilde{\omega})$ be the spectral density of the bath.
Then the total rate $\Gamma_{fi}$ of bath-induced transitions among the
Floquet states $i$ and $f$ of the driven system is expressed as a sum of
partial rates,
\begin{equation}
	\Gamma_{fi} = \sum_{\ell\in{\mathbbm Z}} \, \Gamma_{fi}^{(\ell)} \; ,
\label{eq:TOR}
\end{equation}
where
\begin{equation}
	\Gamma_{fi}^{(\ell)} = 2\pi \, | V_{fi}^{(\ell)} |^2 \,
	N(\omega_{fi}^{(\ell)})  \, J(|\omega_{fi}^{(\ell)}|) \; .
\label{eq:GFI}
\end{equation}
The evaluation of this formula requires a definite specification of the
quasienergy representatives for each state when computing the transition
frequencies~(\ref{eq:OFI}); this speci\-fication also fixes the
representatives of the Floquet functions which enter the matrix
elements~(\ref{eq:FDV}). An alternative choice of representatives would lead
to a shift of the Fourier index~$\ell$, but leaves the sum~(\ref{eq:TOR})
invariant.

These total rates~(\ref{eq:TOR}) now determine the desired quasi\-stationary
distribution~$\{ p_n \}$ as a solution to the equation~\cite{BreuerEtAl00}
\begin{equation}
	\sum_m \big( \Gamma_{nm} p_m - \Gamma_{mn} p_n \big) = 0 \; .
\label{eq:PME}
\end{equation}	
It deserves to be emphasized again that the very details of the system-bath
coupling enter here, so that the precise form of the respective distribution
$\{ p_n \}$ may depend strongly on such details~\cite{Kohn01}.

\subsection{The lift from $s = 1/2$ to general $s$}
\label{sec:2b}

We will make heavy use of a procedure which allows one to transfer a solution
to the Schr\"odinger equation for a spin with spin quantum number $s = 1/2$
in a time-dependent external field to a solution of the corresponding
Schr\"odinger equation for general~$s$, see also Ref.~\cite{Schmidt18}.
This procedure essentially rests on the fact that a spin-$s$ state can
be represented as a direct symmetrized product of $2s$ spin-$1/2$ states,
as exposed by Landau and Lifshitz~\cite{LaLiQM81}. It does not appear
to be widely known, but has been applied already in 1987 to the coherent
evolution of a laser-driven $N$-level system possessing an $SU(2)$ dynamic
symmetry~\cite{Hioe87}, and more recently to the spin-$s$ Landau-Zener
problem~\cite{PokrovskySinitsyn04}. Here we briefly sketch this method.

Let $t \mapsto \Psi(t)\in SU(2)$ be a smooth curve such that
$\Psi(0) = {\mathbbm 1}$, as given by a $2 \times 2$-matrix of the form
\begin{equation}
	\Psi(t) = \left( \begin{array}{rr}
		z_1(t) & z_2(t) \\
		-z_2^\ast(t) & z_1^\ast(t) \end{array} \right)
\end{equation}
with complex functions $z_1(t)$, $z_2(t)$ obeying
$|z_1(t)|^2 + |z_2(t)|^2 = 1$ for all times~$t$. One then has
\begin{equation}
  	\left(\frac{\rd}{\rd t} \Psi(t) \right) \, \Psi(t)^{-1}
	\equiv -\ri H(t) \in su(2) \; ,
\label{eq:LIE}
\end{equation}
where $su(2)$ denotes the Lie algebra of $SU(2)$, {\em i.e.\/}, the space
of anti-Hermitean, traceless $2\times 2$-matrices which is closed under
commutation~\cite{Hall15}. Hence the columns $|\psi_1(t)\rangle$,
$|\psi_2(t)\rangle$ of $\Psi(t)$ are linearly independent solutions of the
Schr\"odinger equation
\begin{equation}
 	\ri \frac{\rd}{\rd t} |\psi_j(t)\rangle = H(t) \, | \psi_j(t) \rangle
	\; , \quad j = 1,2 \; .
\end{equation}
Next we consider the well-known irreducible Lie algebra representation
of $su(2)$,
\begin{equation}
  	r^{(s)} : su(2) \longrightarrow su(2s+1) \; ,
\end{equation}
which is parametrized by a spin quantum number $s$ such that
$2s\in{\mathbbm N}$, together with the corresponding irreducible
group representation (``irrep'' for brevity)
\begin{equation}
  	R^{(s)} : SU(2) \longrightarrow SU(2s+1) \; .
\end{equation}	
One then has~\cite{Hall15}
\begin{equation}
  	r^{(s)}(\ri s_j) = \ri S_j \; ,
	\quad j = x,y,z \; ,
\label{eq:REP}
\end{equation}
where $s_j = \sigma_j/2$ denote the three $s = 1/2$ spin operators given by
the Pauli matrices $\sigma_j$, and the $S_j$ denote the corresponding spin
operators for general~$s$. Recall the standard matrices
\begin{eqnarray}
  	\left( S_z \right)_{m,n} & = & n \, \delta_{mn} \; ,
\\
   	\left( S_x \right)_{m,n} & = & \left\{ \begin{array}{r@{\quad:\quad}l}
	\phantom{\pm} \displaystyle \frac{1}{2}
	\sqrt{s(s+1) -n(n \pm 1)} & m = n \pm 1 \; , \\
	0 & \mbox{else} \; , \end{array} \right.
\nonumber \\	
 	\left( S_y \right)_{m,n} & = & \left\{ \begin{array}{r@{\quad:\quad}l}
	\pm \displaystyle \frac{1}{2\ri}
	\sqrt{s(s+1) - n(n\pm 1)} & m = n \pm 1 \; , \\
	0 & \mbox{else} \; ,	\end{array} \right.
\nonumber
\end{eqnarray}
where $m,n = s, s-1, \ldots, -s$, and
\begin{equation}
 	S_\pm \equiv S_x \pm \ri S_y \; .
\end{equation}
It follows from the general theory of representations~\cite{Hall15} that
$r^{(s)}$ and $R^{(s)}$ may be applied to Eq.~(\ref{eq:LIE}) and yield
\begin{equation}
 	\left( \frac{\rd}{\rd t} R^{(s)} \, \Psi(t) \right) \,
	\left(  R^{(s)} \, \Psi(t) \right)^{-1} =
  	r^{(s)}\big( - \ri H(t) \big) \; .
\label{eq:LSG}
\end{equation}
Since the traceless matrix $H(t)$ can always be written as a linear combination
of the spin operators $s_j$ it acquires the form of a Zeeman term with a
time-dependent magnetic field ${\mathbf b}(t)$, namely,
\begin{equation}
 	H(t) ={\mathbf b}(t) \cdot {\mathbf s}
	=\sum_{j=1}^{3} b_j(t) \, s_j \; ,
\label{eq:HZF}
\end{equation}
and Eq.~(\ref{eq:REP}) now implies
\begin{equation}
  	r^{(s)}\big(-\ri H(t) \big)
	= -\ri \, {\mathbf b}(t) \cdot {\mathbf S}
	= -\ri \sum_{j=1}^{3} b_j(t) \, S_j \; .
\end{equation}
Hence, the ``lifted'' matrix
\begin{equation}
 	\Psi^{(s)}(t) \equiv R^{(s)} \big( \Psi(t) \big)
\label{eq:LIF}
\end{equation}
will be a matrix solution to the lifted Schr\"odinger equation
\begin{equation}
	\ri \frac{\rd}{\rd t} \Psi^{(s)}(t) =
	{\mathbf b}(t)\cdot {\mathbf S} \, \Psi^{(s)}(t) \; .
\label{eq:LSE}
\end{equation}
Note that the matrix $\Psi^{(s)}(t)$ is unitary, and hence its columns span
the general $(2s+1)$-dimensional solution space of the lifted Schr\"odinger
equation~(\ref{eq:LSE}). The decisive step of this procedure, namely, the
lift from Eq.~(\ref{eq:LIE}) to Eq.~(\ref{eq:LSG}), is further illustrated
in Appendix~\ref{sec:APA} with the help of an elementary example.

\section{Floquet formulation of the Rabi problem}
\label{sec:III}

\subsection{Floquet decomposition for $s = 1/2$}
\label{sec:3a}

A spin $1/2$ subjected to both a constant magnetic field applied in the
$z$-direction and an orthogonal, circularly polarized time-periodic field,
as constituting the classic Rabi problem~\cite{Rabi37}, is described by the
Hamiltonian
\begin{equation}
	H(t) = \frac{\omega_0}{2} \sigma_z
	+ \frac{F}{2}(\sigma_x \cos \omega t + \sigma_y \sin \omega t) \; .
\label{eq:TLH}
\end{equation}
Here $\omega_0 > 0$ denotes the transition frequency pertaining to the spin
states in the static field alone, while~$F$, carrying the dimension of a
frequency in our system of units, denotes the amplitude of the periodic drive.
This is a special form of the Zeeman Hamiltonian~(\ref{eq:HZF}) with the
particular choices
\begin{eqnarray}
   	b_x(t) & = &  F \cos\omega t
\nonumber \\
  	b_y(t) & = & F \sin\omega t
\nonumber \\
   	b_z(t) & = & \omega_0 \; .
\label{eq:MAF}
\end{eqnarray}
The Floquet states~(\ref{eq:FST}) brought about by this
Hamiltonian~(\ref{eq:TLH}) are given by~\cite{HolthausJust94}
\begin{equation}
	| \psi_\pm(t) \rangle =
	\frac{\re^{\mp \ri \Omega t/2}}{\sqrt{2\Omega}} \left(
	\begin{array}{r}
	\pm \sqrt{\Omega \pm \delta} \; \re^{-\ri\omega t/2} \\
            \sqrt{\Omega \mp \delta} \; \re^{+\ri\omega t/2}
	\end{array} \right) \; ,
\label{eq:FCP}
\end{equation}
where
\begin{equation}
	\delta = \omega_0 - \omega
\end{equation}
denotes the detuning of the transition frequency~$\omega_0$ from the
driving frequency~$\omega$, and $\Omega$ is the Rabi frequency,
\begin{equation}
	\Omega = \sqrt{\delta^2 + F^2} \; .
\label{eq:RAB}
\end{equation}	
The $2 \times 2$-matrix $\Psi(t)$ constructed from these states does not
satisfy $\Psi(0) = {\mathbbm 1}$. This is of no concern, since $\Psi(t)$ could
be replaced by $\Psi(t) \big( \Psi(0) \big)^{-1}$. The distinct advantage of
these Floquet solutions~(\ref{eq:FCP}) lies in the fact that they yield a
particularly convenient starting point for the lifting procedure outlined in
Sec.~\ref{sec:2b}: One has
\begin{eqnarray}
	\Psi(t) & = & \frac{1}{\sqrt{2\Omega}} \left( \begin{array}{rr}
	\sqrt{\Omega + \delta} &
       -\sqrt{\Omega - \delta} \\
	\re^{\ri\omega t}\sqrt{\Omega - \delta} &
	\re^{\ri\omega t}\sqrt{\Omega + \delta}
	\end{array} \right)
\nonumber \\	& & \times
	\left( \begin{array}{cc}
	\re^{-\ri(\omega + \Omega)t/2} & 0 \\
	0 & \re^{-\ri(\omega - \Omega)t/2}
	\end{array} \right)
\nonumber \\	& \equiv &
	P(t) \, \re^{-\ri\omega t/2} \, \exp(-\ri\Omega t \, s_z) \; .
\phantom{\int_a^b}			
\label{eq:FDC}
\end{eqnarray}
This decomposition possesses the general Floquet form
\begin{equation}
	\Psi(t) = P(t) \exp(-\ri G t) \; ,
\label{eq:GFF}
\end{equation}
where the unitary matrix $P(t) = P(t+T)$ again is $T$-periodic, and the
eigenvalues of the ``Floquet matrix''~$G$, to be obtained from the matrix
logarithm of $\Psi(T) \big( \Psi(0) \big)^{-1} = \exp(-\ri G T)$, provide the
system's quasienergies~\cite{Shirley65,Salzman74,GesztesyMitter81,Holthaus16}.
Since~$G$ already is diagonal in this representation~(\ref{eq:FDC}), the
quasienergies of a spin $1/2$ driven by a circularly polarized field
according to the Hamiltonian~(\ref{eq:TLH}) can be read off immediately:
\begin{equation}
 	\varepsilon_\pm = \frac{\omega\pm \Omega}{2}
	\quad \bmod \omega \; ,
\end{equation}
satisfying $\varepsilon_+ + \varepsilon_- = 0 \, \bmod \,\omega$.
For later application we express the periodic part $P(t)$ of the
decomposition~(\ref{eq:FDC}) in the following way:
\begin{eqnarray}
	P(t) & = & \re^{\ri\omega t/2} \left( \begin{array}{cc}
	\re^{-\ri\omega t/2} & 0 \\
	0 & \re^{+\ri\omega t/2} \end{array} \right)
\nonumber \\	& & \times
	\frac{1}{\sqrt{2\Omega}} \left( \begin{array}{rr}
	\sqrt{\Omega + \delta} & -\sqrt{\Omega - \delta} \\
	\sqrt{\Omega - \delta} &  \sqrt{\Omega + \delta}
	\end{array} \right)
\nonumber \\	& \equiv &
	\re^{\ri \omega t/2} \exp(-\ri \omega t \, s_z) \, \Xi \; .
\phantom{\int_a^b}
\label{eq:RPM}
\end{eqnarray}			
The time-independent matrix~$\Xi = \Psi(0)$ introduced here can be written as
\begin{equation}
	\Xi = \exp\left( -\ri\lambda \, s_y \right) =
 	\left( \begin{array}{rr}
 	\cos(\lambda/2) & -\sin(\lambda/2) \\
 	\sin(\lambda/2) &  \cos(\lambda/2)
	\end{array} \right)
\label{eq:DFM}
\end{equation}
with
\begin{equation}
 	\lambda/2 = \arccos\left(
	\sqrt{\frac{\Omega + \delta}{2\Omega}} \right) \; .
\end{equation}
Hence, one has the identities
\begin{eqnarray}
 	\Xi^\dagger s_x \Xi & = &
	\frac{\delta}{\Omega} \, s_x +
	\frac{\sqrt{\Omega^2 - \delta^2}}{\Omega } \, s_z
\nonumber \\	
 	\Xi^\dagger s_y \Xi & = & s_y
\phantom{\frac{\delta}{\Omega}} \; ,
\nonumber \\
 	\Xi^\dagger s_z \Xi & = &
	\frac{\delta}{\Omega} \, s_z -
	\frac{\sqrt{\Omega^2 - \delta^2}}{\Omega } \, s_x \; ,
\label{eq:MSM}	
\end{eqnarray}
which will be put into use in both Sec.~\ref{sec:IV} and Sec.~\ref{sec:V}.

\subsection{Floquet decomposition for general $s$}
\label{sec:3b}

Replacing the spin-$1/2$ operators $s_j = \sigma_j/2$ in the
Hamiltonian~(\ref{eq:TLH}) by their conterparts~$S_j$ for general spin
quantum number~$s$, one obtains
\begin{eqnarray}
	H^{(s)}(t) & = & {\mathbf b}(t) \cdot {\mathbf S}
\nonumber \\	& = &	
	\omega_0\,S_z + F \left(S_x\,\cos\omega t+S_y\,\sin\omega t\right)
 	\; .
\label{eq:HGS}
\end{eqnarray}
According to Sec.~\ref{sec:2b} the general matrix solution to the corresponding
Schr\"odinger equation~(\ref{eq:LSE}) now is obtained as the lift~(\ref{eq:LIF})
of the $2 \times 2$-matrix~(\ref{eq:FDC}). Invoking Eqs.~(\ref{eq:RPM}) and
(\ref{eq:DFM}), and applying the irrep $R^{(s)}$ to this decomposition yields
\begin{eqnarray}
	\Psi^{(s)}(t) & = & R^{(s)}\!\Big(
	\exp(-\ri \omega t \, s_z)
	\exp\left( -\ri\lambda \, s_y \right)
	\exp(-\ri\Omega t \, s_z) \Big)
\nonumber \\	& = &
	\exp(-\ri \omega t \, S_z)
	\exp\left( -\ri\lambda \, S_y \right)
	\exp(-\ri\Omega t \, S_z) \; .
\phantom{\int_a^b}		
\end{eqnarray}
In order to bring this factorization into the standard Floquet form analogous
to Eq.~(\ref{eq:GFF}),
\begin{equation}
	\Psi^{(s)}(t) = P^{(s)}(t) \exp\!\big(-\ri G^{(s)} t\big)	
\end{equation}
with a $T$-periodic matrix $P^{(s)}(t) = P^{(s)}(t+T)$, we have to
distinguish two cases:
\begin{description}
\item[{\em (i)\/}] For {\em integer\/} $s$ we may set
\begin{eqnarray}
	P^{(s)}(t) & = & \exp(-\ri \omega t \, S_z) \,
	\exp\left( -\ri\lambda \, S_y \right) \; ,
\nonumber \\
	G^{(s)} & = & \Omega \, S_z \; .
\label{eq:PSI}	
\end{eqnarray}

\item[{\em (ii)\/}] For {\em half-integer\/} $s$ the requirement that
$P^{(s)}(t)$ be $T$-periodic demands insertion of additional factors
$\re^{\pm\ri\omega t/2}$, in analogy to the representation~(\ref{eq:RPM}) for
$s = 1/2$. This gives
\begin{eqnarray}
	P^{(s)}(t) & = & \re^{\ri\omega t/2}
	\exp(-\ri \omega t \, S_z) \,
	\exp\left( -\ri\lambda \, S_y \right) \; ,
\nonumber \\
	G^{(s)} & = & \frac{\omega}{2}{\mathbbm 1}^{(s)} + \Omega\,S_z \; ,
\label{eq:PSH}
\end{eqnarray}
where ${\mathbbm 1}^{(s)}$ indicates the unit matrix in ${\mathbbm C}^{2s+1}$.

\end{description}

Denoting the eigenstates of $S_z$ as $| m \rangle$, such that
$S_z | m \rangle = m | m \rangle$, we now introduce Floquet functions and
their quasienergies according to the prescription
\begin{eqnarray}
	\Psi^{(s)}(t) \, | m \rangle & = &
	P^{(s)}(t) \, | m \rangle \,
	\exp\!\left(-\ri \varepsilon_m t\right)
\nonumber \\	& \equiv &
	| u_m(t) \rangle \, \exp\!\left(-\ri \varepsilon_m t\right) \; ,
\end{eqnarray}
implying the particular choice
\begin{equation}
	|u_m(t) \rangle = P^{(s)}(t) \, | m \rangle
\label{eq:TPF}
\end{equation}	
of $T$-periodic Floquet functions. The associated quasi\-energy representatives
then are
\begin{equation}
	\varepsilon_m = m \Omega	
\label{eq:QEI}
\end{equation}
for integer~$s$ according to case {\em (i)\/}, or
\begin{equation}
	\varepsilon_m = \frac{\omega}{2} + m \Omega
\label{eq:QEH}
\end{equation}
for half-integer~$s$ according to case {\em (ii)\/}, with $m = -s, \ldots, s$.
This convenient choice of representatives will be presupposed in the following
for computing the partial rates~(\ref{eq:GFI}). Observe that there is a
further, physically important distinction to be made at this point: When
the driving amplitude vanishes, that is, for $F \to 0$, the Rabi
frequency~(\ref{eq:RAB}) does not reduce to the detuning, but rather to the
absolute value of the detuning, $\Omega \to | \omega_0 - \omega |$. Hence,
for $F \to 0$ the Floquet functions~(\ref{eq:TPF}) ``connect'' to
$| m \rangle$ only for red detuning, when $\omega < \omega_0$, but to
$| -m \rangle$ for blue detuning, when $ \omega> \omega_0$. Hence, under
blue detuning the labeling of the Floquet functions and their quasienergies
effectively is reversed with respect to the eigenstates of $S_z$. This feature
needs to be kept in mind for correctly assessing the following results.

We also note that in the adiabatic low-frequency limit, when the spin is
exposed to an arbitrarily slowly varying magnetic field enabling adiabatic
following to the instantaneous energy eigenstates, the quasienergies should
be given by the one-cyle averages of the instantaneous energy eigenvalues.
Indeed, in this limit the Rabi frequency~(\ref{eq:RAB}) reduces to
$\Omega = \sqrt{\omega_0^2 + F^2}$, while the time-independent instantaneous
energy levels are $E_m = m\,\Omega$, yielding the expected identity.

\section{The quasistationary distribution}
\label{sec:IV}

Now we stipulate that the periodically driven spin be coupled to a thermal
bath of harmonic oscillators, as sketched in Sec.~\ref{sec:2a}, taking the
coupling operator to be of the simple form~\cite{ExplainV}
\begin{equation}
	V = \gamma \, S_x \; .
\label{eq:VGS}
\end{equation}
In order to calculate the Fourier decompositions~(\ref{eq:FDV}) of the Floquet
matrix elements of~$V$, and referring to the above representation~(\ref{eq:TPF})
of the Floquet functions, we thus need to consider the operator
\begin{eqnarray}
	& &
 	P^{(s)\dagger}(t) \, S_x \, P^{(s)}(t)
\nonumber \\	& = &	
	\exp\left(\ri\lambda \, S_y \right)	
  	\exp\left(\ri\omega t \, S_z \right) \, S_x\,
	\exp\left(- \ri\omega t \, S_z \right)
	\exp\left(- \ri\lambda \, S_y \right)
\nonumber \\	& \equiv &
	\sum_{\ell\in{\mathbbm Z}} \, V^{(\ell)}
	\exp(\ri\ell\omega t) \; ;
 \end{eqnarray}
note that the additional phase factor $\re^{\ri\omega t/2}$ contained in the
expression~(\ref{eq:PSH}) for $P^{(s)}(t)$ with half-integer~$s$ cancels here.
Using the $su(2)$ commutation relations and their counterparts for general~$s$,
we deduce
\begin{eqnarray}
	& &
	\exp\left(\ri\omega t \, S_z \right) \, S_x\,
	\exp\left(- \ri\omega t \, S_z \right)
\nonumber \\	& = &	
	\frac{1}{2}\left( \re^{\ri\omega t} \, S_+
	+  \re^{-\ri\omega t} \, S_- \right) \; .
 \end{eqnarray}
Hence, as in the case $s = 1/2$ studied in Ref.~\cite{LangemeyerHolthaus14},
the only non-vanishing Fourier components $V^{(\ell)}$ occur for $\ell=\pm 1$:
\begin{equation}
 	V^{(\pm 1)}= \frac{\gamma}{2}
	\exp\left(\ri\lambda \, S_y \right) S_\pm
	\exp\left(- \ri\lambda \, S_y \right) \; .
\end{equation}	
Applying $R^{(s)}$ to Eqs.~(\ref{eq:MSM}), this yields
\begin{equation}
 	V^{(\pm 1)} =  \frac{\gamma}{2}\left(\frac{\delta}{\Omega} \, S_x \,
	+\frac{\sqrt{\Omega^2-\delta^2}}{\Omega} \, S_z\,
	\pm\ri\, S_y \right) \; .
\label{eq:VPM}
\end{equation}
Thus, $V^{(\pm 1)}$ is a tridiagonal matrix. For computing the partial
transition rates~(\ref{eq:GFI}) we therefore have to consider only frequencies
$\omega_{mn}^{(\pm1)}$ of pseudotransitions~\cite{LangemeyerHolthaus14}, for
which $m = n$, and of transitions between neighboring Floquet states,
$m = n \pm 1$. For evaluating the definition~(\ref{eq:OFI}) one now has to
resort to the quasienergy representatives~(\ref{eq:QEI}) and~(\ref{eq:QEH})
which belong to the Floquet functions~(\ref{eq:TPF}) entering here, giving
\begin{equation}
 	\omega_{mn}^{(\pm 1)}= \left\{\begin{array}{r@{\quad:\quad}l}
 			\pm \omega 	& m = n   \;, \\
    \phantom{-} \Omega  \pm \omega 	& m = n+1 \;, \\
	       -\Omega 	\pm \omega	& m = n-1 \; .
  	\end{array} \right.
\label{eq:OMN}
\end{equation}
According to the Pauli master equation~(\ref{eq:PME}), the quasi\-stationary
distribution $\{ p_m \}_{m = s,\ldots,-s}$ which establishes itself under the
combined influence of time-periodic driving and the thermal oscillator bath
is the eigenvector of a tridiagonal matrix $\widetilde{\Gamma}$ corresponding
to the eigenvalue $0$, where $\widetilde{\Gamma}$ is obtained from
$\Gamma\equiv\Gamma^{(1)}+\Gamma^{(-1)}$ by subtracting from the diagonal
elements the respective column sums, {\em i.e.\/},
\begin{equation}
	\widetilde{\Gamma}_{mn} = \Gamma_{mn}
	- \delta_{mn} \sum_{k=-s}^{s}\Gamma_{kn} \; .
\label{eq:GAT}
\end{equation}	
Since $\widetilde{\Gamma}$ is tridiagonal with non-vanishing secondary
diagonal elements, this eigenvector is unique up to normalization. Moreover,
it is evident that we only need the matrix elements of~$\Gamma$ in the
secondary diagonals for calculating the quasistationary distribution,
whereas the diagonal elements will be required for computing the dissipation
rate~\cite{LangemeyerHolthaus14}.

The very fact that $V^{(\pm 1)}$, and hence $\Gamma$, merely is a tridiagonal
matrix has a conceptually important consequence: It enforces detailed
balance, meaning that each term of the sum~(\ref{eq:PME}) vanishes
individually. With $\Gamma$ being tridiagonal, this sum reduces to
\begin{eqnarray}
	( \Gamma_{n,n-1} \, p_{n-1} - \Gamma_{n-1,n} \, p_n ) & &
\nonumber \\	
      + \; ( \Gamma_{n,n+1} \, p_{n+1} - \Gamma_{n+1,n} \, p_n ) & = & 0
\label{eq:TTL}
\end{eqnarray}
for all $n = -s+1$, $\ldots\,$, $s-1$, since the term with $m = n$ in
Eq.~(\ref{eq:PME}) cancels. In the border cases $n = -s$ or $n = s$ this
identity still holds, but only one bracket survives. Upon setting the first
bracket in this Eq.~(\ref{eq:TTL}) to zero, one obtains
\begin{equation}
	\frac{p_n}{p_{n-1}} = \frac{\Gamma_{n,n-1}}{\Gamma_{n-1,n}}
\end{equation}
for $n = -s+1$, $\ldots\,$, $s-1$. Together with the normalization requirement,
this relation alone already determines the entire distribution $\{ p_m \}$. In
particular, it entails
\begin{equation}
	\frac{p_{n+1}}{p_n} = \frac{\Gamma_{n+1,n}}{\Gamma_{n,n+1}} \; ,
\label{eq:PDP}
\end{equation}
thus ensuring that also the second bracket in Eq.~(\ref{eq:TTL}) vanishes,
confirming detailed balance.

Now one needs to observe that the sign of the transition
frequencies~(\ref{eq:OMN}) depends on the relative magnitude of the driving
frequency~$\omega$ and the Rabi frequency~$\Omega$; recall that the
distinction between positive and negative transition frequencies ---
physically corresponding to the distinction between annihilation and creation
of bath phonons --- leads to the two different expressions~(\ref{eq:NOP})
and~(\ref{eq:NOM}) entering the transition rates~(\ref{eq:GFI}). This prompts
us to distinguish between the low-frequency case $0 < \omega < \Omega$ and the
high-frequency case $0 < \Omega < \omega$ in the following. The resonant case
$\omega = \Omega$ constitutes a special problem; this is best dealt with by
taking the appropriate limits of the results obtained in the other two cases.

Finally, a further factor of substantial importance is the spectral density
$J(\widetilde{\omega})$, which may allow one to manipulate the quasistationary
distribution to a considerable extent~\cite{DiermannEtAl19}. For the sake
of simplicity and transparent discussion, here we assume that
$J(\widetilde{\omega}) \equiv J_0$ is constant.

\subsection{Low-frequency case $0 < \omega < \Omega$}
\label{sec:4a}

In order to utilize the above Eq.~(\ref{eq:PDP}) for determining the
distribution $\{ p_m \}$ recursively we only need to evaluate the partial
rates $\Gamma_{m,m+1}^{(\pm 1)}$ and $\Gamma_{m+1,m}^{(\pm 1)}$
according to the general prescription~(\ref{eq:GFI}), making use of the
particular representation~(\ref{eq:VPM}). In the low-frequency case this
leads to the expressions
\begin{eqnarray}
   	\Gamma^{(\pm 1)}_{m,m+1} & = &  \frac{s(s+1)-m(m+1)}
	{16\left(1 - \re^{-\beta(\Omega\mp\omega)} \right)}
	\left(\frac{\Omega\mp\delta}{\Omega}\right)^2
\nonumber \\	
  	\Gamma^{(\pm 1)}_{m+1,m} & = & \frac{s(s+1)-m(m+1)}
	{16\left(\re^{\beta(\Omega\pm\omega)} - 1 \right)}
	\left(\frac{\Omega\pm\delta}{\Omega}\right)^2
\label{eq:GAL}
\end{eqnarray}
which have been scaled by $\Gamma_0 = 2\pi\gamma^2 J_0$, and have thus been
made dimensionless. Evidently, these representations imply that the desired
ratio
\begin{equation}
	\frac{\Gamma_{m+1,m}}{\Gamma_{m,m+1}} \equiv q_L
\end{equation}	
is independent of both~$s$ and~$m$; a tedious but straightforward calculation
readily yields 	
\begin{equation}
    q_L = \frac{ 2\delta\Omega \sinh(\beta\omega)
    + \left( \delta^2 + \Omega^2 \right)
    \left(\re^{-\beta\Omega} - \cosh(\beta\omega) \right)}
                    { 2\delta\Omega \sinh(\beta\omega)
    - \left( \delta^2 + \Omega^2 \right)
    \left(\re^{ \beta\Omega} - \cosh(\beta\omega) \right)} \; .
\label{eq:DQL}
\end{equation}
Therefore, the quasistationary occupation probabilities of the Floquet states
can be written in the form
\begin{equation}
	p_m = \frac{1}{Z_L} q_L^m
\label{eq:PML}
\end{equation}
with $m = -s, \ldots, s$, and $Z_L$ ensuring normalization. Hence, not only
does one find detailed balance here, but the occupation probabilities even
generate a finite geometric sequence.

\subsection{High-frequency case $0 < \Omega < \omega$}
\label{sec:4b}

Analogously, in the high-frequency case $0 < \Omega < \omega$ we require
the dimensionless partial rates
\begin{eqnarray}
   	\Gamma^{(\pm 1)}_{m,m+1} & = & \pm \frac{s(s+1) - m(m+1)}
	{16\left(\re^{\beta(\pm\omega - \Omega )} - 1 \right)}
	\left(\frac{\Omega\mp\delta}{\Omega}\right)^2
\nonumber \\
  	\Gamma^{(\pm 1)}_{m+1,m} & = &\pm \frac{s(s+1) - m(m+1)}
	{16\left(\re^{\beta(\Omega \pm \omega)} - 1 \right)}
	\left(\frac{\Omega\pm\delta}{\Omega}\right)^2
\label{eq:GAH}
\end{eqnarray}
for constructing the matrix $\Gamma = \Gamma^{(1)} + \Gamma^{(-1)}$.
Once more, the ratio
\begin{equation}
	\frac{\Gamma_{m+1,m}}{\Gamma_{m,m+1}} \equiv q_H
\end{equation}	
does depend neither on~$s$ nor on~$m$, so that the occupation probabilities
again form a geometric sequence; after some juggling, one finds
\begin{equation}
   	q_H = \frac{ \left( \delta^2 + \Omega^2 \right)
	\sinh(\beta\omega) + 2 \delta\Omega
        \left(\re^{-\beta\Omega} - \cosh(\beta\omega) \right)}
	                { \left( \delta^2 + \Omega^2 \right)
	\sinh(\beta\omega) - 2 \delta\Omega
	\left(\re^{ \beta\Omega } - \cosh(\beta\omega) \right)} \; .
\label{eq:DQH}
\end{equation}
Thus, the high-frequency Floquet-state occupation probabilities are
given by
\begin{equation}
	p_m = \frac{1}{Z_H} q_H^m
\label{eq:PMH}	
\end{equation}	
analogously to Eq.~(\ref{eq:PML}).

\subsection{The quasitemperature}
\label{sec:4c}

The fact that, on the one hand, the distribution of occupation probabilities
$\{ p_m \}$ is a geometric one in both the low- and the high frequency case,
and that the quasienergy representatives $\varepsilon_m$ of all Floquet
states can be taken to be equidistant on the other, suggests to write this
distribution in Boltzmann form,
\begin{equation}
	p_m = \frac{1}{Z} \exp(-\vartheta\varepsilon_m) \; ,
\label{eq:GDI}
\end{equation}
where $Z$ is adjusted such that this distribution is normalized,
$\sum_{m=-s}^s p_m = 1$. Evidently, the parameter $\vartheta$ introduced here
plays the role of an inverse {\em quasitemperature\/}, and $Z$ is a formal
analog of a canonical partition function~\cite{Pathria11}. We emphasize that
we are dealing with a nonequilibrium steady state of the driven system which
does not possess a temperature in the sense of equilibrium thermodynamics.
However, this nonequilibrium steady state is characterized by a Boltzmannian
distribution~(\ref{eq:GDI}) into which a single parameter enters {\em as if\/}
it were a temperature; hence, the designation ``quasi\-temperature'' is well
justified. Needless to say, although the driven system is in contact with a
bath which possesses a given temperature, its quasitemperature can be quite
different from that temperature.

Evidently, the definition of such a quasitemperature still involves a certain
degree of arbitrariness, as it requires to single out a specific representative
of the quasienergy class of each Floquet state. In principle, one could employ
the representatives
\begin{equation}
	\varepsilon_m = m(\Omega + r\omega)
\label{eq:CQR}
\end{equation}
with arbitrary fixed $r \in {\mathbbm Z}$ for integer~$s$, and odd
$r \in {\mathbbm Z}$ for half-integer~$s$; this freedom can be exploited to
equip the respective quasi\-temperature with
properties deemed to be desirable. For instance, when setting $r = +1$
($r = -1$) for red (blue) detuning, one obtains representatives which for
$F \to 0$ connect continuously to the energy eigenvalues of the undriven spin
described by $H_0 = \omega_0 S_z$, thus guaranteeing that the quasitemperature
introduced via these representatives reduces to the temperature of the ambient
bath in the limit of vanishing driving amplitude.

Here we take a different route, and introduce a quasi\-temperature on the
grounds of the representatives~(\ref{eq:QEI}) and~(\ref{eq:QEH}) selected
earlier for computing the rates~(\ref{eq:GFI}), so that the distribution
acquires the practical form
\begin{equation}
	p_m = \frac{1}{Z_q} \exp(-\vartheta \Omega m)
\end{equation}
for both integer and half-integer~$s$. Introducing, for ease of notation,
\begin{equation}
  	q \equiv \left\{ \begin{array}{r@{\quad:\quad}l}
	q_L	& 0 < \omega < \Omega	\\
	q_H	& 0 < \Omega < \omega 	\; ,
	\end{array} \right.
\label{eq:DIS}
\end{equation}
with $q_L$ and $q_H$ as given by the expressions~(\ref{eq:DQL}) and
(\ref{eq:DQH}), respectively, this leads immediately to the definition
\begin{equation}
  	\vartheta \equiv -\frac{\ln q}{\Omega} \; .
\label{eq:DQT}
\end{equation}
Since $q$, being a ratio of two rates, may adopt any value between zero and
$+\infty$, the quasitemperature then ranges from $-\infty$ to $+\infty$.
Notwithstanding the various possible definitions of a quasitemperature,
the very distribution $\{ p_m \}$ itself, governing the observable physics,
is determined uniquely. Also observe that, when $F \to 0$,
\begin{equation}
	q \to \left\{ \begin{array}{r@{\quad:\quad}l}
	\re^{-\beta\omega_0} & \omega < \omega_0 \\
	\re^{+\beta\omega_0} & \omega > \omega_0 \; .
	\end{array} \right.	
\label{eq:UPS}
\end{equation}
Hence, in the absence of the time-periodic driving field the above
solutions~(\ref{eq:PML}) and~(\ref{eq:PMH}) correctly lead to a Boltzmann
distribution with the inverse bath temperature~$\beta$, thereby indicating
thermal equilibrium of the spin system; the unfamiliar ``plus''-sign appearing
in this limit~(\ref{eq:UPS}) in the case of blue detuning merely reflects
the reversed labeling already discussed in the paragraph following
Eq.~(\ref{eq:QEH}).

The dimensionless inverse quasitemperature $\omega_0\vartheta$ ultimately
depends on the scaled driving amplitude~$F/\omega_0$, the scaled driving
frequency~$\omega/\omega_0$, and the dimensionless inverse actual
temperature~$\omega_0 \beta$ of the heat bath, but not on the spin quantum
number~$s$. In contrast, the partition function $Z_q$ depends on~$s$:
\begin{equation}
 	Z_q = \sum_{m=-s}^s \exp(-\vartheta\Omega m) =
	\frac{\sinh\!\left(\displaystyle\frac{2s+1}{2}\,\vartheta\Omega\right)}
	     {\sinh\!\left(\displaystyle\frac{\vartheta\Omega}{2}\right)} \; .
\label{eq:QPF}
\end{equation}

\begin{figure}[t]
\centering
\includegraphics[width=1.0\linewidth]{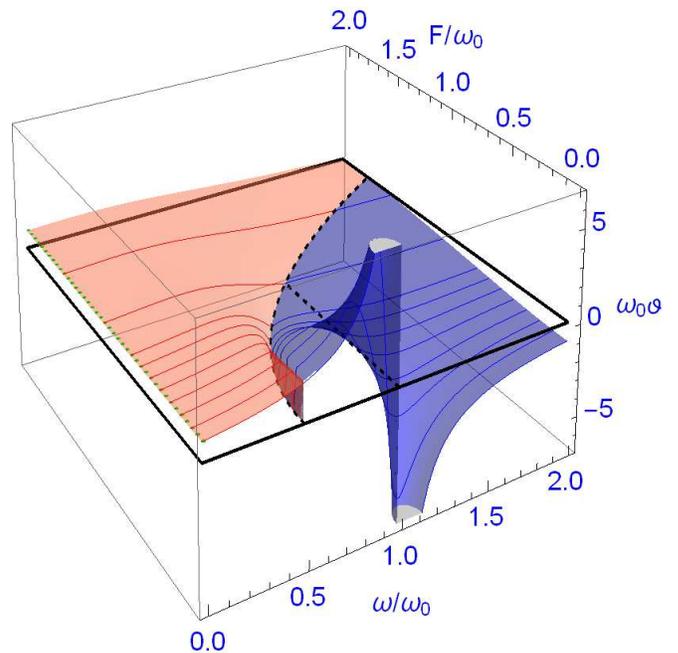}
\caption{(Color online) Dimensionless inverse quasitemperature $\omega_0\vartheta$ of the
	spin system under the influence of a circularly polarized monochromatic
	driving force with amplitude~$F$ and frequency~$\omega$, being
	coupled to a harmonic-oscillator bath of inverse actual temperature
	$\omega_0\beta = 1$. The blue (dark) part of the graph corresponds to the
	high-frequency regime $0 < \Omega < \omega$, the red (gray) part to the
	low-frequency regime $0 < \omega < \Omega$. Along the line segment
	$\omega = \omega_0$ with $F < \omega_0$ and along the parabola
	$\omega = \omega_c$ given by Eq.~(\ref{eq:OMC}), both marked by black
	dashes, the inverse quasitemperature vanishes. A few functions
	$\vartheta(\omega)$ for constant~$F$ are highlighted. The limit
	$\omega_0\vartheta = \omega_0\beta$ for $\omega/\omega_0 \to 0$
	is indicated by a green (dotted) line.}
\label{F_1}
\end{figure}

\begin{figure}[t]
\centering
\includegraphics[width=1.0\linewidth]{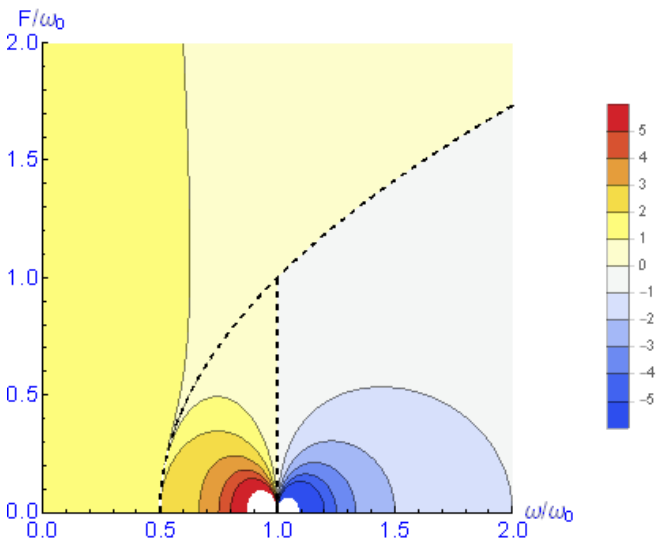}
\caption{(Color online) Contour plot of the dimensionless inverse quasitemperature
	$\omega_0\vartheta$ with the same parameters as in Figure \ref{F_1}.
	The line segment $\omega = \omega_0$ with $F < \omega_0$ and the
	parabola $\omega = \omega_c$, where $\vartheta$ vanishes, are again
	marked by black dashes.}
\label{F_1A}
\end{figure}

The inverse quasitemperature~$\vartheta$ vanishes --- meaning that the
periodically driven system effectively becomes infinitely hot, so that all its
Floquet states are populated equally --- regardless of the bath temperature,
if either $\omega = \omega_0$ while $0 <  F < \omega_0$, or if
\begin{equation}
 	\omega = \omega_c \equiv
	\frac{1}{2\omega_0} \left(F^2+\omega_0^2\right) \; ,
\label{eq:OMC}	
\end{equation}
with the latter equation defining the boundary $\omega = \Omega$ between
the low- and the high-frequency regime, see Figs.~\ref{F_1} and~\ref{F_1A}.
Along this boundary the quasienergies of all Floquet states are degenerate
(modulo~$\omega$), so that the appearance of infinite quasitemperature here
can be understood as a resonance effect. In terms of the inverse
quasitemperatures, denoted $\vartheta_L$ ($\vartheta_H$) in the low-
(high-) frequency regime, there is a continuous change from $\vartheta_L$
to $\vartheta_H$, since
\begin{equation}
  	\lim_{\omega\uparrow   \omega_c} \vartheta_L =
	\lim_{\omega\downarrow \omega_c} \vartheta_H = 0 \; .
\end{equation}
However, the two functions $\vartheta_L$ and $\vartheta_H$ do not join
smoothly at $\omega = \omega_c$, since their derivatives with respect
to~$\omega$ adopt different limits:
\begin{eqnarray}
  	\lim_{\omega\uparrow \omega_c}
	\frac{\partial\vartheta_L}{\partial\omega} & = &
	-\frac{4\beta\omega _0^3 \left(F^4 + \omega _0^4\right)}
	      {F^4 \left(F^2 + \omega_0^2 \right)^2} \; ,
\nonumber \\
  	\lim_{\omega\downarrow \omega_c}
	\frac{\partial\vartheta_H}{\partial\omega} & = &
        \frac{4 \beta  \omega _0^3 \left(F-\omega _0\right)
	                           \left(F+\omega _0\right)}
	     {F^4 \left(F^2+\omega _0^2\right)} \; .
\label{eq:DWO}
\end{eqnarray}

We will now investigate the behavior of the function
$\omega_0\vartheta(\omega_0\beta,\omega/\omega_0,F/\omega_0)$ in the limits
corresponding to the four sideways faces of the box bounding the plot displayed
in Fig.~\ref{F_1}:

{\em (i)\/} As already noted at the end of Sec.~\ref{sec:3b}, in the
low-frequency limit $\omega/\omega_0 \to 0$ the quasienergies~(\ref{eq:QEI})
and~(\ref{eq:QEH}) approach the actual energies
$E_m = m\sqrt{\omega_0^2 + F^2}$ (with $m = s, \ldots, -s$) of a spin exposed
to a slowly varying drive. Hence, in this limit the periodic thermodynamics
investigated here must reduce to the usual thermodynamics described by a
canonical ensemble; in particular, the inverse  quasitemperature~$\vartheta$
must approach the true inverse bath temperature~$\beta$. This expectation is
borne out by the leading term of the low-frequency expansion
\begin{widetext}
\begin{eqnarray}
	\omega_0\vartheta_L
	& = &
	\omega _0\beta
	+ \frac{2\omega_0\beta (\omega/\omega_0)}{(F/\omega_0)^2 + 2}
\nonumber \\	& &
	+ \frac{\omega_0\beta (\omega/\omega_0)^2}
	       {2\big( (F/\omega_0)^2 + 2 \big)^2}
	\left( 8 - 4 (F/\omega _0)^2
	- \frac{ (F/\omega _0)^4 \,\omega_0\beta \,
	\coth \left(\frac{\omega_0\beta}{2}\sqrt{ (F/\omega_0)^2 + 1} \right)}
	       {\sqrt{ (F/\omega_0)^2 + 1}} \right)
	+ O(\omega/\omega_0)^3 \; .
\end{eqnarray}
\end{widetext}

{\em (ii)\/} Next we consider the ultrahigh-frequency limit
$\omega/\omega_0 \to \infty$, keeping both $\omega_0\beta$ and $F/\omega_0$
fixed. Inspecting $q_H$ as defined by Eq.~(\ref{eq:DQH}), and observing that
asymptotically $\Omega = \sqrt{(\omega_0 - \omega)^2 + F^2} \sim \omega - \omega_0$
for $\omega/\omega_0 \to \infty$, one finds $q_H \to \exp(+\beta\omega_0)$.
Recalling the reversed labeling of Floquet states for blue detuning, which
accounts for the ``plus'' sign, this implies that in this limit the
Floquet-state distribution~(\ref{eq:PMH}) equals the Boltzmann distribution
for the undriven energy eigenstates with the bath temperature. This finding can
intuitively be understood as an averaging principle: In the ultrahigh-frequency
limit the effect of the exernal drive on the occupation probabilities averages
out, leaving one with the original thermal distribution. On the other hand,
there is no such averaging principle for quasienergies; the ac Stark shift
(that is, the deviation of the quasienergies from the energy eigenvalues of
the undriven system) increases as $\omega$. Since our quasi\-temperature is
defined with respect to the quasienergies, it follows that the quasitemperature
has to vanish at high frequencies: By virtue of Eq.~(\ref{eq:DQT}) one has
$\vartheta_H \sim -\beta\omega_0/\omega$, and hence
\begin{equation}\label{LIM2}
    \lim_{\omega/\omega_0 \to \infty}
    \omega_0\vartheta_H \; = \; 0 \; ;
\end{equation}
as shown by Fig.~\ref{F_1}, this limit is approached with negative
quasitemperatures.

{\em (iii)} The static limit of vanishing driving amplitude,
$F/\omega_0 \to 0$, yields
\begin{equation}
	\lim_{F/\omega_0 \to 0} \omega_0\vartheta_L
	\; =  	
	\lim_{F/\omega_0 \to 0} \omega_0\vartheta_H
	\; = \;  	
	\frac{\omega_0\beta}{1 - \omega/\omega_0} \; ,
\label{eq:LFN}
\end{equation}
easily deduced from the limits~(\ref{eq:UPS}) in combination with the
definition~(\ref{eq:DQT}). Once again, this seemingly strange expression,
exhibiting a pole at $\omega = \omega_0$ which is prominently visible in
Fig.~\ref{F_1}, is fully in agreement with the requirement that the
periodic thermodynamics should reduce to ordinary thermodynamics when the
time-periodic driving force vanishes. Namely, for $F/\omega_0 \to 0$
the system possesses the energy eigenvalues $E_m = m\omega_0$, whereas
the quasienergy representatives~(\ref{eq:QEI}) or~(\ref{eq:QEH}) reduce
to $\varepsilon_m = m |\omega_0 - \omega|$ or
$\varepsilon_m = \omega/2 + m|\omega_0 - \omega|$ with $m = s,\ldots,-s$,
while the Floquet states approach the energy eigenstates of
$H_0 = \omega_0 S_z$. In this limit the parametrization of their occupation
probabilities in terms of either the quasithermal distribution~(\ref{eq:GDI})
or the standard canonical distribution must lead to identical values, implying
that the usual Boltzmann factor $\exp(-\beta m \omega_0)$ must become
proportional to $\exp\left(-\vartheta m |\omega_0 - \omega| \right)$. Hence,
one has $\beta\omega_0 = \vartheta(\omega_0 - \omega)$ for $F/\omega_0 \to 0$
and $0 < \omega < \omega_0$, which immediately furnishes the above
expression~(\ref{eq:LFN}). In the case of blue detuning, for
$0 < \omega_0 < \omega$, the reversed labeling of the quasienergies yields
an additional ``minus'' sign, again leading to the limit~(\ref{eq:LFN}).

\begin{figure}[t]
\centering
\includegraphics[width=1.0\linewidth]{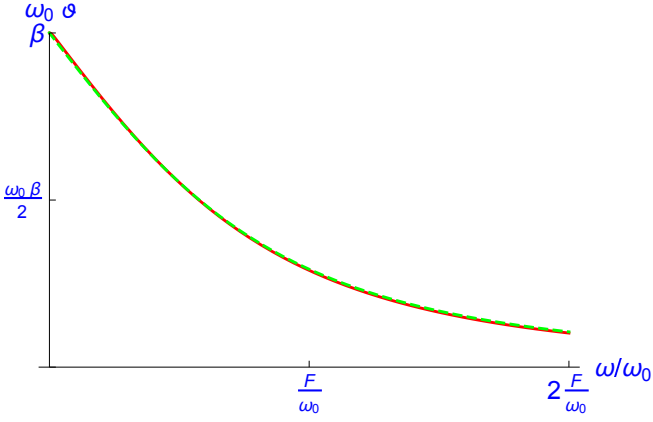}
\caption{(Color online) Inverse dimensionless quasitemperature $\omega_0\vartheta$ for
	$\omega_0\beta = 1$ and $F/\omega_0 = 100$ as a function of
	$\omega/\omega_0$ in the regime $0 < \omega/\omega_0 < 2F/\omega_0$.
	We show the exact values of $\omega_0\vartheta_L$ (red line),
	as well as the asymptotic form~(\ref{eq:CLI}) (green dashes).}
\label{F_2}
\end{figure}

{\em (iv)\/} In the converse strong-driving limit $F/\omega_0 \to \infty$ we
first focus on the regime $\omega \sim F$ where $\vartheta = \vartheta_L > 0$.
After some transformations we obtain the asymptotic form
\begin{equation}
 	\omega_0\vartheta_L \sim \omega_0\beta
	-\frac{\omega\beta}{\sqrt{ (F/\omega_0)^2 + (\omega/\omega_0)^2}}
\label{eq:CLI}
\end{equation}
in this regime, so that the inverse quasitemperature decreases monotonically
with increasing driving amplitude, as exemplified by Fig.~\ref{F_2}.

\begin{figure}[t]
\centering
\includegraphics[width=1.0\linewidth]{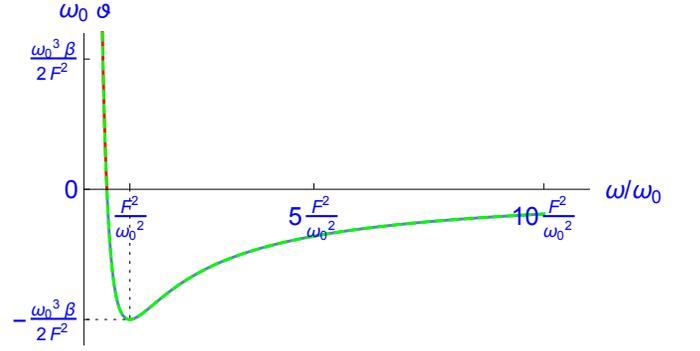}
\caption{(Color online) Inverse dimensionless quasitemperature $\omega_0\vartheta$ for
	$\omega_0\beta = 1$ and $F/\omega_0 = 100$ as a function of
	$\omega/\omega_0$ in the regime $\omega/\omega_0 > (F/\omega_0)^2/2$.
  	We show the exact values of $\omega_0\vartheta_L$ (red solid line) and
  	$\omega_0\vartheta_H$ (blue solid line), together with the asymptotic
	form~(\ref{eq:AFM}) (green dashes).}
\label{F_3}
\end{figure}

In contrast, for $\omega > \omega_c$ as given by Eq.~(\ref{eq:OMC}) we have
$\vartheta = \vartheta_H < 0$. Asymptotically, here we find
\begin{equation}
	\omega_0\vartheta_H \sim
	- \frac{\omega_0\beta}{\omega/\omega_0}
	+ \frac{\omega_0\beta}{2(\omega/\omega_0)^2} \, (F/\omega_0)^2 \; .
\label{eq:AFM}
\end{equation}

This asymptotic function exhibits a pronounced minimum at $\omega/\omega_0 =
(F/\omega_0)^2$ of depth $\omega_0\vartheta_H \sim -\omega_0^3 \beta/(2F^2)$,
depicted in Fig.~\ref{F_3}. This minimum can be understood as a result of
two opposing trends: On the one hand, the averaging principle requires that
the occupation probabilities approach the ``undriven'' Boltzmann distribution
with the bath temperature for large $\omega/\omega_0$, which means that the
inverse quasitemperature tends to a finite value proportional to $-\beta$
for large, but finite driving frequencies. On the other hand, the factor
$1/\Omega$ in Eq.~(\ref{eq:DQT}) becomes more and more predominant and forces
$\vartheta$ to approach~$0$ in the high-frequency regime.

\section{Application: Periodically driven paramagnets}
\label{sec:V}

\subsection{Calculation of quasithermal expectation values}
\label{sec:5a}

Starting from the proposition that the Floquet states of a periodically driven
spin system be populated according to the distribution~(\ref{eq:GDI}) we
introduce the quasithermal average of the spin component in the direction
of the static field,
\begin{equation}
	\langle  S_z \rangle_q \equiv \frac{1}{Z} \sum_{m=-s}^{s}
	\langle u_m(t)|S_z|u_m(t)\rangle \,
	\exp(-\vartheta\varepsilon_m) \; .
\label{eq:QTA}
\end{equation}
In order to evaluate this expression we utilize the
representation~(\ref{eq:TPF}) of the Floquet functions, giving
\begin{eqnarray}
	Z \langle S_z \rangle_q & = & \sum_{m=-s}^{s}
	\langle m | P^{(s)\dagger} \, S_z \, P^{(s)} | m \rangle \,
	\exp(-\vartheta\varepsilon_m)
\nonumber \\	& = &	
	\mbox{Tr}\left( P^{(s)\dagger} \, S_z \, P^{(s)} \,
	\exp\!\big(-\vartheta G^{(s)} \big) \right) \; .
\label{eq:TRE}
\end{eqnarray}
Next, we resort once more to the lifting technique: For $s = 1/2$,
the decomposition~(\ref{eq:RPM}) readily yields
\begin{eqnarray}
	P^\dagger s_z P & = & \Xi^\dagger s_z \Xi
\nonumber \\	& = &
	\frac{\delta}{\Omega } \, s_z
	-\frac{\sqrt{\Omega ^2 - \delta ^2}}{\Omega } \, s_x
\end{eqnarray}
by virtue of Eq.~(\ref{eq:MSM}).
Applying the irrep~$r^{(s)}$, we deduce
\begin{eqnarray}
	P^{(s)\dagger} \, S_z \, P^{(s)} & = &
	r^{(s)}\big( P^\dagger s_z P \big)	
\nonumber \\	& = &
	\frac{\delta}{\Omega } \, S_z
	-\frac{\sqrt{\Omega ^2 - \delta ^2}}{\Omega } \, S_x \; .
\end{eqnarray}
Inserting this into the above identity~(\ref{eq:TRE}), and calculating the
trace in the eigenbasis of~$S_z$, we obtain the important result
\begin{equation}
 	\langle S_z \rangle_q =
	\frac{1}{Z_q} \frac{\omega_0 - \omega}{\Omega} \,
	\sum_{m=-s}^{s} m \, \re^{-\vartheta\Omega \, m} \; ,
\label{eq:ESZ}
\end{equation}
valid for both integer and half-integer~$s$. In particular, this shows that the
quasithermal expectation value~(\ref{eq:QTA}) does not depend on time, despite
the time-dependence of the Floquet functions. Although the unusual-looking
prefactor $(\omega_0 - \omega)/\Omega$ indeed implies that the $z$-component
of the magnetization vanishes for $\omega = \omega_0$, it does not necessarily
imply that the magnetization reverses its direction when $\omega$ is varied
across $\omega_0$, since the reversal of the prefector's sign can be
compensated by a simultaneous change of the sign of the quasitemperature,
as it happens for low driving amplitudes according to Eq.~(\ref{eq:LFN}).

\subsection{Response of paramagnetic materials to circularly polarized
            driving fields}
\label{sec:5b}

As an experimentally accessible example of the above considerations, and thus
as a possible laboratory application of periodic thermodynamics, we consider
the magnetization of an ideal paramagnetic substance under the influence of
both a static magnetic field applied in the $z$-direction, and a circularly
polarized oscillating magnetic field applied in the $x$-$y$-plane. In order
to facilitate comparison with the literature, here we re-install the Planck
constant~$\hbar$ and the Boltzmann constant~$\kB$.

We assume that the magnetic atoms of the substance have an electron shell with
total angular momentum~$J$, resulting from the coupling of orbital angular
momentum and spin, giving the magnetic moment $\mu = g_J \mu_{\rm B} J$.
Here $\mu_{\rm B}$ denotes the Bohr magneton, and $g_J$ is the Land\'e
$g$-factor which may assume both signs; for the sake of definiteness, here
we assume $g_J < 0$. In the presence of a constant magnetic field $B_0$
this moment gives rise to the energy levels
\begin{equation}
	E_m = -m \, g_J \mu_{\rm B} B_0 \equiv m\,\hbar\omega_0 \; ,
\label{eq:ELP}
\end{equation}
where $m = -J \ldots, J$ is the magnetic quantum number. Hence, with $g_J < 0$
the spin tends to align antiparallel to the applied magnetic field, favoring
$m = -J$.

Let us briefly recall the usual textbook treatment of the ensuing thermal
paramagnetism within the canonical ensemble~\cite{FowlerGuggenheim39,Pathria11},
assuming the substance to possess a temperature~$T$. Then the canonical
partition function
\begin{eqnarray}
 	Z_0 & = &
	\sum_{m=-J}^J \exp\!\left(-\frac{E_m}{\kB T}\right)=
	\sum_{m=-J}^{J} \exp\!\left({-}m\frac{\hbar\omega_0}{k_B T}\right)
\nonumber \\	& = &	
	\frac{\sinh\!\left(\displaystyle\frac{2J+1}{2J}\,y_0\right)}
	     {\sinh\!\left(\displaystyle\frac{y_0}{2J}\right)}
\label{eq:PFM}
\end{eqnarray}
which depends on the dimensionless quantity
\begin{equation}
  	y_0  \equiv -\frac{g_J\mu_{\rm B}B_0}{\kB T} \, J
	= \frac{\hbar\omega_0}{\kB T}\, J
\end{equation}
serves as moment-generating function, in the sense that the thermal expectation
value of the magnetization $M$ is obtained by taking the appropriate derivative
of its logarithm, namely,
\begin{eqnarray}
	\langle  M \rangle & = &{\frac{N}{V} \,\langle  \mu \rangle}=
         \frac{N}{V} \, g_J \mu_{\rm B} \langle m \rangle
\nonumber \\	& = &		
	 \frac{N}{V}\frac{\partial}{\partial B_0}\,k_B T \ln Z_0 \; .
\label{eq:MAG}
\end{eqnarray}
Here $N/V$ denotes the density of contributing atoms. Working out this
prescription, one finds the magnetization~\cite{FowlerGuggenheim39,Pathria11}
\begin{equation}
	\langle M \rangle = M_0 \, B_J(y_0) \; ,
\label{eq:THP}
\end{equation}
where
\begin{equation}
  	M_0 =- \frac{N}{V }\, g_J\mu_{\rm B} J
\label{eq:SMA}	
\end{equation}
denotes the saturation magnetization, and
\begin{equation}
	B_J(y) \equiv \frac{2J+1}{2J}\coth\!\left( \frac{2J+1}{2J}y\right)
	- \frac{1}{2J}\coth\!\left(\frac{y}{2J}\right)
\end{equation}
is the so-called Brillouin function of order~$J$~\cite{Brillouin27}; this
theoretical prediction~(\ref{eq:THP}) has been beautifully confirmed in
low-temperature experiments with paramagnetic ions by Henry~\cite{Henry52}
already in 1952. In the weak-field limit $\mu_{\rm B}B_0 \ll \kB T$ one may
use to approximation
\begin{equation}
	B_J(y) \approx \frac{J+1}{J} \frac{y}{3}
	\quad \mbox{for} \quad | y | \ll 1 \; ,
\label{eq:ABJ}
\end{equation}
giving
\begin{equation}
	\langle M \rangle \approx  \frac{N}{V} \,
	\frac{(g_J\mu_{\rm B})^2 J(J+1)}{3\kB T} \, B_0 \; .
\label{eq:WFM}
\end{equation}	

Returning to periodic thermodynamics, let us add the circularly polarized
field $B_1(\cos\omega t,\sin\omega t,0)$ perpendicularly to the constant one.
Then the Rabi frequency (\ref{eq:RAB}) can be written as
\begin{equation}
 	\Omega = \sqrt{(\omega_0 - \omega)^2 + (g_J\mu_{\rm B}B_1/\hbar)^2}
	\; ,
\label{eq:RAQ}
\end{equation}
where
\begin{equation}
  	\omega_0 = -\frac{g_J\mu_{\rm B}B_0}{\hbar} \;
\end{equation}
measures the strength of the static field in accordance with
Eq.~(\ref{eq:ELP}). Assuming that the spins' environment is correctly
described by a thermal oscillator bath with constant spectral density~$J_0$,
and the coupling to that environment is given by the expression~(\ref{eq:VGS}),
the Floquet-state occupation probabilities are governed by the
distribution~(\ref{eq:GDI}), and we can invoke the above result~(\ref{eq:ESZ})
to write the observable magnetization in the form
\begin{eqnarray}
\label{eq:QMA}
	\langle M \rangle_q & = &
	-\frac{N}{V} \, g_J \mu_{\rm B} \langle S_z \rangle_q
\\	& = &
	-\frac{N}{V} \, \frac{g_J\mu_{\rm B}}{Z_q} \,
	\frac{\omega_0 - \omega}{\Omega} \sum_{m=-J}^J m \,
	\exp\!\left({-}m\frac{\hbar\Omega}{\kB\tau}\right) \; ,
\nonumber 		
\end{eqnarray}
where $\tau = 1/(\kB\vartheta)$ is the quasitemperature, and
\begin{equation}
	Z_q = \sum_{m=-J}^{J}
	\exp\!\left({-}m\frac{\hbar\Omega}{\kB \tau}\right)
\end{equation}
is the corresponding partition function~(\ref{eq:QPF}). Quite remarkably,
this expression~(\ref{eq:QMA}) is a perfect formal analog of the previous
Eq.~(\ref{eq:MAG}), since we have
\begin{equation}
	\langle M \rangle_q = \frac{N}{V} \,
	\frac{\partial}{\partial B_0} \, \kB \tau \ln Z_q \; ,
\label{eq:DPF}
\end{equation}
taking into account the nonlinear dependence of the Rabi
frequency~(\ref{eq:RAQ}) on the static field strength~$B_0$. Hence, the
resulting quasithermal magnetization can be expressed in a manner analogous
to Eq.~(\ref{eq:THP}), namely,
\begin{equation}
	\langle M \rangle_q = M_1 \, B_J(y_1) \; ,
\label{eq:MMB}
\end{equation}
with modified saturation magnetization
\begin{equation}
 	M_1 = \frac{\omega_0 - \omega}{\Omega } \, M_0 \; ,
 \label{eq:M1}
\end{equation}
and the argument of the Brillouin function now depending on the
quasitemperature,
\begin{equation}
 	y_1 = \frac{\hbar\Omega}{\kB\tau} \, J \; .
\end{equation}	
For consistency, this prediction~(\ref{eq:MMB}) must reduce to the usual
weak-field magnetiziation~(\ref{eq:WFM}) when both $B_1 \to 0$ and
$\mu_{\rm B} B_0 \ll \kB T$. This is ensured by the limit~(\ref{eq:LFN}):
For sufficiently small~$B_1$, the quasitemperature~$\tau$ is related to the
actual bath temperature~$T$ through
\begin{equation}
	\frac{1}{\kB \tau} \approx \frac{\omega_0}{\omega_0 - \omega} \,
	\frac{1}{\kB T} \; .
\end{equation}	
Inserting this into Eq.~(\ref{eq:MMB}), one can employ the
approximation~(\ref{eq:ABJ}) for frequencies~$\omega$ not too close to
$\omega_0$. In this way one recovers the expected expression~(\ref{eq:WFM})
unless $\omega \approx \omega_0$, in which case one has
$\langle M \rangle_q \approx 0$.

\begin{figure}[t]
\centering
\includegraphics[width=1.0\linewidth]{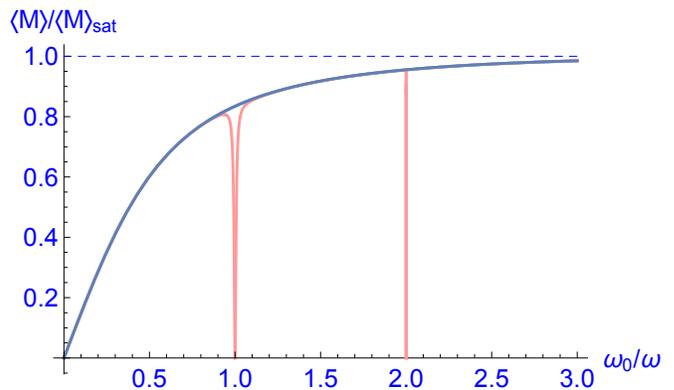}
\caption{(Color online) Magnetizations $\langle M \rangle$ divided by their saturation
	magnetizations as functions of the scaled strength $\omega_0/\omega$
	of the static field. The blue (dark) curve represents the ordinary thermal
	magnetization~(\ref{eq:THP}), the red (gray) one the quasi\-thermal
	magnetization~(\ref{eq:MMB}) for weak driving,
	$g_J\mu_{\rm B}B_1/(\hbar\omega) = 0.01$. Here we have set
	$\kB T = \hbar\omega$ and $J = 7/2$.}
\label{F_4}
\end{figure}

As is evident from the above discussion, under typical conditions of electron
spin resonance (ESR) with weak driving amplitudes, such that $B_1/B_0$ is on
the order of $10^{-2}$ or less~\cite{EatonEtAl10}, the difference between the
quasithermal magnetization~$\langle M \rangle_q$ and the customary thermal
magnetization~$\langle M \rangle$ is more or less negligible, except for
driving frequencies close to resonance. This is illustrated in Fig.~\ref{F_4}
for $g_J\mu_{\rm B}B_1/(\hbar\omega) = 0.01$, where the bath temperature is
chosen such that $\kB T = \hbar\omega$. Here we have plotted both the usual
magnetization~(\ref{eq:THP}) of an undriven spin system and the quasithermal
magnetization~(\ref{eq:MMB}) of its weakly driven counterpart, normalized
to their respective saturation value, vs.\ the ratio~$\omega_0/\omega$,
representing the scaled strength~$B_0$ of the static field. The vanishing
of the quasi\-thermal magnetization at both $\omega_0/\omega = 1$ and
$\omega_0/\omega = 2$ clearly reflects the appearance of infinite
quasitemperature at these ratios, as already observed in Fig.~\ref{F_1}.

\begin{figure}[t]
\centering
\includegraphics[width=1.0\linewidth]{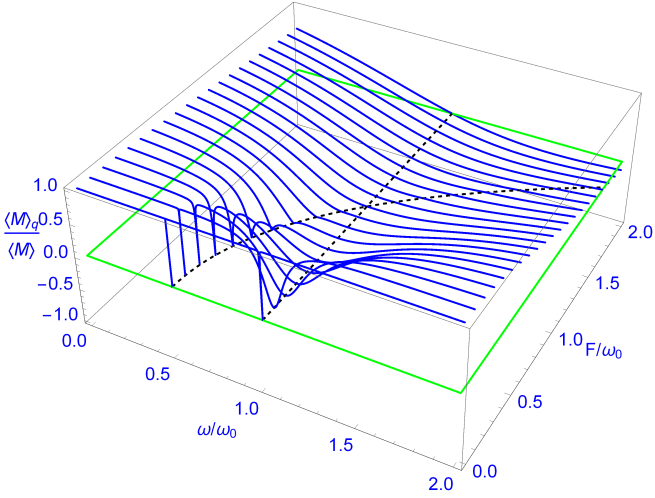}
\caption{(Color online) Ratio $\langle M \rangle_q / \langle M \rangle$ of the
	quasithermal magnetization~(\ref{eq:MMB}) and the customary
	magnetization~(\ref{eq:THP}) as function of $\omega/\omega_0$ and
	$F/\omega_0$, with $F = g_J \mu_{\rm B} B_1/\hbar$. Parameters chosen
	here are $\kB T = \hbar\omega_0$ and $J = 1$. Along the dashed black line
	$\omega = \omega_0$ and along the dashed black parabola $\omega = \omega_c$
	given by Eq.~(\ref{eq:OMC}) the quasi\-thermal magnetization vanishes,
	so that $\langle M \rangle_q / \langle M \rangle$ becomes negative
	for strong driving with frequencies $\omega_0 < \omega < \omega_c$.}
\label{F_5}
\end{figure}

\begin{figure}[t]
\centering
\includegraphics[width=1.0\linewidth]{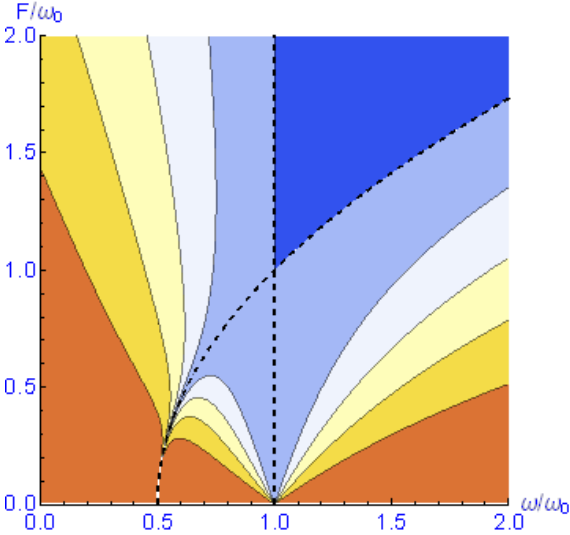}
\caption{(Color online) Contour plot of the ratio $\langle M \rangle_q / \langle M \rangle$
	with the same parameters as in Fig.~\ref{F_5}. The domain in which
	$\langle M \rangle_q / \langle M \rangle$ becomes negative is indicated
	by the dark blue color.}
\label{F_5A}
\end{figure}

In marked contrast, novel types of behavior with  measurable consequences
occur in the regime of non-perturbatively strong driving. A particularly
striking example is provided by Fig.~\ref{F_5}, where
$\kB T = \hbar\omega_0$ and $J = 1$: Under strong driving, the ratio
$\langle M \rangle_q / \langle M \rangle$ actually becomes {\em negative\/}
for frequencies $\omega_0 < \omega < \omega_c$, implying that the paramagnetic
material effectively becomes a diamagnetic one, reflecting the fact that under
strong driving the distribution of Floquet-state occupation probabilities can
differ substantially from the original thermal Boltzmann distribution. This
possibility of turning a para\-magnet into a diamagnet through the application
of strong time-periodic forcing, as further elucidated in Fig.~\ref{F_5A},
is a ``hard'' prediction of periodic thermodynamics which now awaits its
experimental verification.

\section{Dissipation}
\label{sec:VI}

Since a bath-induced transition from a Floquet state~$n$ to a Floquet state~$m$
is accompanied by all frequencies $\omega_{mn}^{(\ell)}$ as introduced in
Eq.~(\ref{eq:OFI}), the rate of energy dissipated in the quasistationary
state is given by~\cite{LangemeyerHolthaus14}
\begin{equation}
	R = -\sum_{mn\ell}
	\omega_{mn}^{(\ell)} \, \Gamma_{mn}^{(\ell)} \, p_n \; .
\label{eq:DIR}
\end{equation}
For consistency it needs to be shown that $R > 0$, so that in the
nonequilibrium steady state characterized by the Floquet-state distribution
$\{ p_n \}$ the energy flows from the driven system into the bath, regardless
of the system's quasitemperature~\cite{DiermannEtAl19}. While this intuitive
expression yields the mean dissipation rate, it is, in principle, also
possible to to obtain the full probability distribution of energy exchanges
between a periodically driven quantum system and a thermalized heat reservoir
by applying the methods developed in Ref.~\cite{GasparinettiEtAl14}.

The expression~(\ref{eq:DIR}) can now be evaluated for all spin quantum
numbers~$s$. In addition to the partial transition rates~(\ref{eq:GFI}) for
neigboring Floquet states $m = n \pm 1$ listed in Secs.~\ref{sec:4a} and
\ref{sec:4b}, Eq.~(\ref{eq:DIR}) also requires the rates for pseudotransitions
with $m = n$. Again dividing by $\Gamma_0 = 2\pi\gamma^2 J_0$, we obtain the
dimensionless diagonal transition rates
\begin{equation}
 	\Gamma_{mm}^{(\pm 1)} =
	\pm \frac{(2 m)^2 F^2}{16 \left(\re^{\pm \beta\omega} - 1\right)
   	\Omega ^2}
\end{equation}
for $m = s, \ldots, -s$, valid for both cases $0 < \omega < \Omega$ and
$0 < \Omega < \omega$. In order to represent~$R$ in a condensed fashion
we define the polynomials
\begin{eqnarray}
  	P_s(q) & \equiv & -2 \sum _{m=0}^{2 s} (m-s)^2 q^m
\nonumber \\	
  	Q_s(q) & \equiv & \frac{1}{2} \sum _{m=0}^{2s - 1} (m+1)(2s - m) q^m
\nonumber \\
  	z_s(q) & \equiv & \sum _{m=0}^{2 s} q^m \; ,
\end{eqnarray}
together with the expression
\begin{equation}
 	A^\pm(q) \equiv
	\frac{\left(\re^{\beta(\pm\omega + \Omega)}q - 1 \right)
	(\delta \pm \Omega )^2 (\omega \pm \Omega )}
 	{\re^{\beta(\pm\omega + \Omega)} - 1} \; .
\end{equation}
Dividing by $\omega_0\Gamma_0$, one obtains a dimensionless dissipation rate
which can now be written in the form
\begin{equation}
	R = \frac{P_s(q) \, \omega \, F^2 + Q_s(q) \,
	\big(A^+(q)  \mp A^-(q) \big)}{8 \, z_s(q) \, \Omega^2} \; ,
\end{equation}	
where one has to insert either $q_L$ or $q_H$ for $q$, in accordance with
the case distinction~(\ref{eq:DIS}), and the ``$\mp$''-sign in the numerator
becomes ``minus'' for $0 < \omega < \Omega$, but ``plus'' for
$0 < \Omega < \omega$.

After resolving all symbols $R$ will be a function of five arguments,
$R = R(s,\beta,\omega,\omega_0,F)$, which makes the discussion more
difficult than in the case of $s = 1/2$ that has been considered in
Ref.~\cite{LangemeyerHolthaus14}. Thus, here we mention only the most
perspicuous aspects of the dissipation function.

\begin{figure}[t]
\centering
\includegraphics[width=1.0\linewidth]{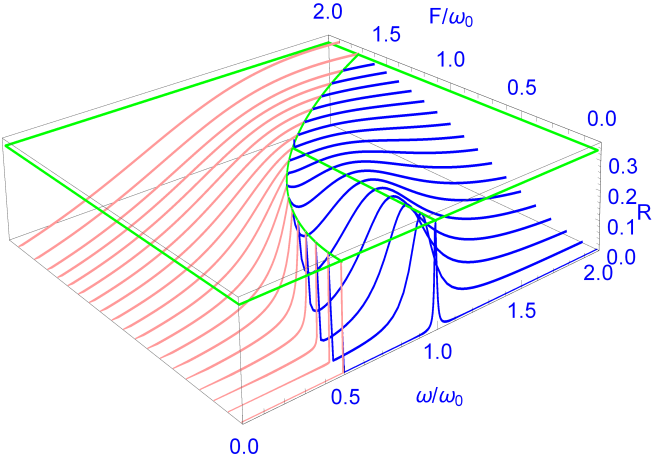}
\caption{(Color online) The dimensionless dissipation rate~$R$ for $s = 1$ and
	$\omega_0\beta = 1$ as a function of $\omega/\omega_0$ and
	$F/\omega_0$. The blue (dark) part of the graph corresponds to the
	high-frequency regime $0 < \Omega < \omega$, the red (gray) one to the
	low-frequency regime $0 < \omega < \Omega$. Along the line
	$\omega = \omega_0$ with $0 < F <\omega_0$ and along the parabola
	$\omega = \omega_c$ given by Eq.~(\ref{eq:OMC}) the dissipation rate
	takes on the constant value $s(s+1)/6 = 1/3$ (green solid lines and parabola).}
\label{F_6}
\end{figure}

\begin{figure}[t]
\centering
\includegraphics[width=1.0\linewidth]{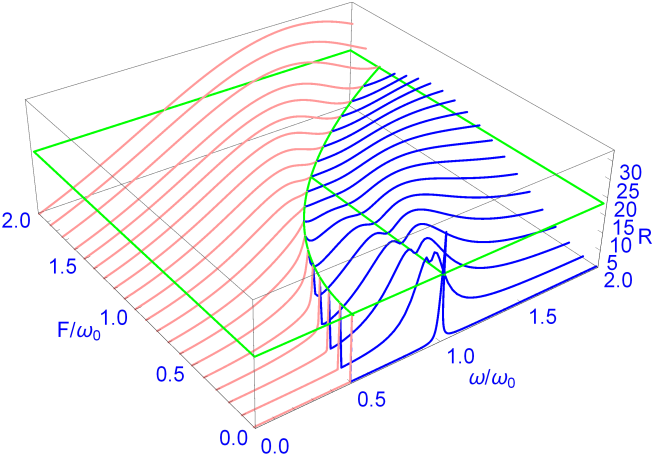}
\caption{(Color online) As Fig.~\ref{F_6}, but for $s = 10$. Along the line
	$\omega = \omega_0$ with $0 < F <\omega_0$ and along the parabola
	$\omega = \omega_c$ given by Eq.~(\ref{eq:OMC}) the dissipation rate
	takes on the constant value $s(s+1)/6 = 55/3$ (green solid lines and parabola).
    The blue (dark)
	curve with $F/\omega_0 \approx 0$ possesses two unresolved sharp
	maxima close to an equally sharp minimum at $\omega {\approx}\omega_0$.}	
\label{F_7}
\end{figure}

Recall that for both $\omega = \omega_0$ with $0 < F <\omega_0$ and
$\omega = \omega_c = \left(F^2 + \omega_0^2\right)/(2\omega_0)$ we have
$q = 1$, and hence $\vartheta = 0$. It turns out that along these two curves
in the $(\omega, \, F)$--plane the dimensionless rate~$R$ takes on the value
\begin{equation}
 	R_0 = \frac{1}{6}s(s+1) \; ,
\end{equation}
as visualized in Figs.~\ref{F_6} and~\ref{F_7} for $s = 1$ and $s = 10$,
respectively. For $1/2 \le s \le 7/2$ this value constitutes a smooth maximum
for $\omega = \omega_0$ and small $F/\omega_0$ which becomes increasingly sharp
for $F/\omega_0 \to 0$. However, for $s > 7/2$ the previous maximum turns into
a local minimum, the sharpness of which increases for $s \to \infty$. In
contrast, along the line $\omega = \omega_c$ this value
$R_0$ remains a local maximum of $R$ for all $s$, as long as
$F < \omega_0$.

\begin{figure}[t]
\centering
\includegraphics[width=1.0\linewidth]{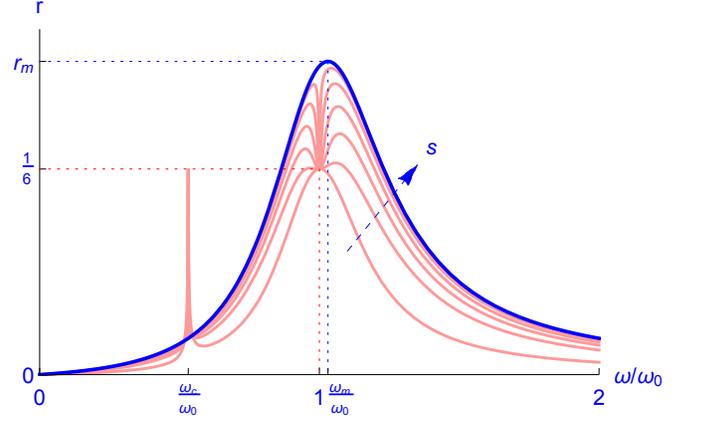}
\caption{(Color online) The scaled dissipation rate $r = R/(s^2 + s)$ for $s = 1/2$, $5$,
	$10$, $20$, $50$, $200$, with bath temperature $\omega_0\beta = 1$
	and driving amplitude $F/\omega_0 = 1/4$, as a function of
	$\omega/\omega_0$. The six red (gray) curves increase with~$s$ for
	$\omega > \omega_c$ as indicated by the dashed arrow. They all meet at the two points with coordinates
	$(\omega_c/\omega_0, 1/6)$ and $(1,1/6)$. The asymptotic
	envelope~(\ref{eq:SDR}) is given by the blue (dark) curve, with its maximum
	$(\omega_m/\omega_0,r_m)$ being determined by~Eqs.~(\ref{eq:IMO})
	and (\ref{eq:IMR}).}
\label{F_8}
\end{figure}

\begin{figure}[t]
\centering
\includegraphics[width=1.0\linewidth]{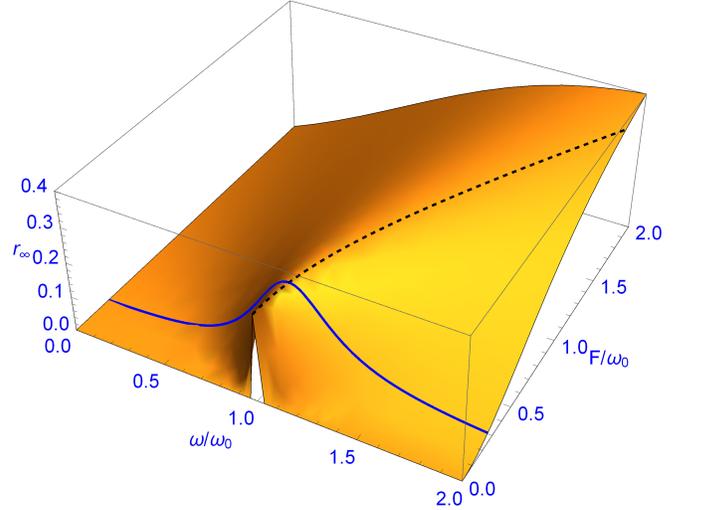}
	\caption{(Color online) The asymptotic limit
	$r_{\infty}(\omega/\omega_0, F/\omega_0)$ of the scaled dissipation
	rate $R/(s^2 + s)$ for $s \to \infty$ according to Eq.~(\ref{eq:SDR}).
	The particular curve with $F/\omega_0 = 1/4$ considered in
	Fig.~\ref{F_8} is shown in blue color (dark solid curve). The maximum values of the
	functions $r_\infty(\omega/\omega_0, F/\omega_0)$ with constant~$F$
	according to Eqs.~(\ref{eq:IMO}) and~(\ref{eq:IMR}) are indicated by
	a black dashed curve.}
\label{F_9}
\end{figure}

For $s\to\infty$ the scaled dissipation rate $r \equiv R/(s^2+s)$
tends to the limit
\begin{equation}
 	r_{\infty}(\omega/\omega_0, F/\omega_0) \equiv
	\frac{\omega}{4\omega_0} \, \frac{F^2 }{F^2+(\omega -\omega_0)^2}
\label{eq:SDR}
\end{equation}
which is independent of the heat-bath temperature. As illustrated by
Fig.~\ref{F_8} the convergence proceeds pointwise except for
$\omega = \omega_c$ and $\omega = \omega_0$ when $F < \omega_0$,
where $r = 1/6$ for all~$s$. For fixed~$F$ the asymptotic function
$r_{\infty}(\omega/\omega_0, F/\omega_0)$ has a global maximum at
\begin{equation}
 	\omega_m/\omega_0 = \sqrt{(F/\omega_0)^2 + 1} \; ,
\label{eq:IMO}	
\end{equation}
adopting the value
\begin{equation}	
	r_m = \frac{1}{8} \left(1 + \sqrt{(F/\omega_0)^2 + 1} \right) \; ,
\label{eq:IMR}
\end{equation}
as depicted in Figs.~\ref{F_8} and~\ref{F_9}. Interestingly, the shape of
$r_\infty$ as a function of $\omega/\omega_0$, indicating a resonance
phenomenon leading to a maximum of the absorbed heat at $\omega = \omega_m$
close to $\omega_0$, is closely related to a prediction based on the
classical Landau-Lifshitz equation~\cite{Gilbert55}, as sketched briefly in
Appendix~\ref{sec:APB}.

\section{Discussion}
\label{sec:VII}

A simple harmonic oscillator which is linearly driven by an external
time-periodic force while kept in contact with a thermal oscillator bath
represents a nonequilibrium system, but nonetheless adopts a steady state
and develops a quasistationary distribution of Floquet-state occupation
probabilities which equals the Boltzmann distribution of the equilibrium model
obtained in the absence of the driving force, being characterized by precisely
the same temperature as that of the bath it is coupled to~\cite{BreuerEtAl00,
LangemeyerHolthaus14}.

The system considered in the present work, a spin with arbitrary spin quantum
number~$s$ exposed to a circularly polarized driving field while interacting
with a bath of thermally occupied harmonic oscillators, may be regarded as the
next basic model in a hierarchy of analytically solvable models on Periodic
Thermodynamics. Exactly as in the case of the linearly forced harmonic
oscillator, the system-bath interaction here induces nearest-neighbor coupling
among the Floquet states of the time-periodically driven system, so that the
model's transition matrix~(\ref{eq:TOR}) is tridiagonal, thus enforcing
detailed balance. Again, the resulting quasistationary Floquet distribution
turns out to be Boltzmannian, but now with a quasitemperature which differs
from the physical temperature of the bath. Already the mere fact that a
time-periodically driven quantum system in its steady state may exhibit a
quasitemperature which is different from the actual temperature of its
environment, and which can be actively controlled by adjusting, {\em e.g.\/},
the amplitude or frequency of the driving force, in itself constitutes a
noteworthy observation, suggesting that periodic thermodynamics generally may
be far more subtle than usual equilibium thermodynamics based on some effective
Floquet Hamiltonian.

For systems exhibiting a geometric distribution $\{ p_n \}$ of Floquet-state
occupation probabilities, the parametrization of the latter in terms of a
quasitemperature is feasible if there are equidistant canonical representatives
of the respective quasienergy classes~\cite{DiermannEtAl19}. Therefore, we
conjecture that the possibility to introduce a quasitemperature remains
restricted to particularly simple integrable systems. However, the concept of
quasistationary Floquet-state occupation probabilities does not require the
introduction of a quasitemperature, and the exploration of the dependence
of the corresponding observable quasistationary expectation values on the
parameters of the driven system and its coupling to its environment constitutes
a major task of periodic thermodynamics in general.

In this sense our model system is not only of basic theoretical interest, but
also leads to novel predictions concerning future experiments with paramagnetic
materials in strong circularly polarized fields. The very existence of a
quasistationary Floquet distribution which is different from the distribution
characterizing thermal equilibrium implies that the magnetic response of
such a periodically driven material can be quite different from that of
the undriven one; as we have demonstrated in Sec.~\ref{sec:5b}, a strong
circularly polarized driving field effectively may turn a paramagnetic
material into a diamagnetic one. While we are not in a position to ascertain
whether the corresponding parameter regime can be reached with already existing
experimental set-ups~\cite{PetukhovEtAl05}, it might be worthwhile to design
specifically targeted measurements for confirming this particularly striking
prediction of periodic thermodynamics.

Yet, there is still more at stake here. When Brillouin published his now-famous
treatise~\cite{Brillouin27} on thermal paramagnetism in 1927, this was
essentially a blueprint for an experimental demonstration of the quantization
of angular momentum, whereas the further thermodynamical input into the
theory was not to be questioned, being backed by the overwhelming generality of
equilibrium thermodynamics~\cite{FowlerGuggenheim39,Pathria11}. At the advent
of periodic thermodynamics more than 90 years later, one faces an inverted
situation: With the quantization of angular momentum being firmly established,
it is nonequilibrium physics in the guise of periodic thermodynamics which is
to be examined in measurements of paramagnetism under time-periodic driving.
As has been stressed already by Kohn~\cite{Kohn01} and clarified by Breuer
{\em et al.\/}~\cite{BreuerEtAl00}, quasistationary Floquet distributions are
not universal, depending on the very form of the system-bath interaction. Here
we have assumed an interaction of the simplistic type~(\ref{eq:HVW})
with coupling~(\ref{eq:VGS}) proportional to the spin operator~$S_x$ on the
system's side and simple creation and annihilation operators~(\ref{eq:PTF})
on the side of the bath, combined with the assumption of a constant spectral
density of the bath, but there are other possibilities. Measurements of
magnetism under strong driving will be sensitive to such issues; two materials
which exhibit precisely the same paramagnetic response in the absence of
time-periodic forcing may react differently to a static magnetic field once
an additional time-periodic field has been added. Thus, despite the formal
similarity of our key results~(\ref{eq:DPF}) and~(\ref{eq:MMB}) to their
historical antecessors~(\ref{eq:MAG}) and~(\ref{eq:THP}), these former
equations may have the potential to open up an altogether new line of research.

\begin{acknowledgments}
This work has been supported by the Deutsche For\-schungsgemeinschaft
(DFG, German Research Foundation) through Projects 355031190, 397122187 and
397300368. We thank all members of the Research Unit FOR~2692 for stimulating
and insightful discussions.
\end{acknowledgments}

\appendix
\section{Remarks on the lifting procedure}
\label{sec:APA}

The technique of lifting a matrix solution of the time-dependent
Schr\"odinger equation for $s = 1/2$ to general~$s$, as reviewed briefly in
Subsec.~\ref{sec:2b}, has been repeatedly used in this paper. In this Appendix
we illustrate this technique with the help of an elementary example from
classical mechanics.

Consider the motion of a rigid body, and let $\vec{r}_0$ denote a constant
position vector in a reference frame fixedly attached to that body
(``B-system''). Then this vector is expressed by
\begin{equation}
	\vec{r}(t) = B(t) \,\vec{ r}_0
\label{eq:AA1}	
\end{equation}
with respect to some inertial laboratory system (``L-system''), where the
rotational $3 \times 3$-matrix $B(t)$ satisfies both $B(t)^\top = B(t)^{-1}$
and $\det B(t) = 1$. Differentiating Eq.~(\ref{eq:AA1}) with respect to time
gives
\begin{equation}
	\frac{\rd}{\rd t} \vec{r}(t) \equiv \dot{\vec{ r}}
	= \dot{B}\,\vec{r}_0 = \left(\dot{B}\,B^\top\right) \vec{r}(t)
	\equiv \Omega(t)\,\vec{r}(t) \; ,
\end{equation}
where the anti-symmetric real $3\times 3$-matrix
\begin{equation}
	\Omega(t) = \dot{B}(t)\,B^\top (t)
\label{eq:AA3}
\end{equation}
has been introduced. This equation $\dot{\vec{ r}} = \Omega(t)\,\vec{r}(t)$
is nothing but the matrix version of the familiar equation
\begin{equation}
	\dot{\vec{ r}} =\vec{\omega} \times \vec{r} \; ,
\end{equation}
involving the vector $\vec{\omega}$ of angular velocity in the L-system; this
vector is constant for all points of the rigid body. Next, let $\vec{\ell}$ be
the angular momentum of the rigid body in the L-system, implying that the
corresponding vectors in the B-system are given by
\begin{eqnarray}
	\vec{\Omega}	& = & B(t)^\top \, \vec{\omega}
\nonumber \\	
	\vec{L}		& = & B(t)^\top \, \vec{\ell} \; .
\label{eq:AA5}
\end{eqnarray}

As is well known, the linear relation between $\vec{\Omega}$ and $\vec{L}$
can be written as
\begin{equation}
	\vec{L} =\Theta \, \vec{\Omega} \; ,
\end{equation}
where the time-independent, symmetric $3\times 3$-matrix~$\Theta$ denotes
the inertial tensor of the rigid body in the B-system. Exploiting
Eqs.~(\ref{eq:AA5}), one obtains the corresponding inertial tensor $\theta(t)$
in the L-system,
\begin{equation}
	\vec{\ell} = B\,\vec{L} = B \, \Theta \, \vec{\Omega}
	=\left( B \, \Theta \, B^\top \right) \vec{\omega}
	\equiv \theta(t)\,\vec{\omega} \; .
\label{eq:AA7}
\end{equation}
Utilizing $ \dot{B} B^\top + B {\dot B}^\top = 0$ and, hence,
\begin{eqnarray}
	{\dot B}^\top = -B^\top \dot{B} B^\top = -B^\top \Omega \; ,	
\end{eqnarray}
its time derivative takes the form
\begin{eqnarray}
	\dot{\theta} & = &
	\phantom{\Omega} \dot{B} \Theta B^\top + B \Theta \dot{B}^\top
\nonumber \\	& = &	
	\Omega B \Theta B^\top + B \Theta (-B^\top \Omega)
	\; =\; \left[\Omega,\theta\right] \; ,
\label{eq:AA9}
\end{eqnarray}
with the last bracket denoting the commutator of two $3\times 3$-matrices.

While this equation~(\ref{eq:AA9}) has been derived above in an elementary
though somewhat tedious manner, we will now show that it follows directly
from Eq.~(\ref{eq:AA3}) by group-theoretical arguments.

To this end, we translate the above considerations into the appropriate
group-theoretical language. Obviously, the matrices $B(t)$ connecting the
B- and the L-system are elements of the Lie group $SO(3)$, and $\Omega(t)$
belongs to the associated Lie algebra $so(3)$ of real anti-symmetric
$3\times 3$-matrices. Its defining equation~(\ref{eq:AA3}) then is recognized
as an immediate analog of the matrix Schr\"odinger equation~(\ref{eq:LIE})
which has served as starting point in Subsec.~\ref{sec:2b}. Moreover, the
transformation $\Theta \mapsto \theta = B\,\Theta\,B^\top$ implied by
Eq.~(\ref{eq:AA7}) can be viewed as an operation of $B$ on the space of
symmetric $3\times 3$-matrices~$\Theta$, and hence as a $6$-dimensional
representation $R$ of $SO(3)$:
\begin{equation}
	\theta \; = \; R(B)(\Theta) \; = \; B\,\Theta\,B^\top \; .
\label{eq:AA10}	
\end{equation}	
It is not irreducible but can be split into $1$- and $5$-dimensional irreps
due to the rotational invariance of the trace of $\Theta$. The corresponding
$6$-dimensional representation $r$ of the Lie algebra $so(3)$ is then given by
\begin{equation}
  	r(\Omega)(\Theta) = \left[ \Omega , \Theta \right] \; .
\end{equation}
This follows if $B$ is written as $B = {\mathbbm 1} + \epsilon\Omega$ with
an anti-symmetric $3 \times 3$-matrix $\Omega$, so that Eq.~(\ref{eq:AA10})
yields
\begin{eqnarray}
	R({\mathbbm 1} + \epsilon\Omega)(\Theta) & = &
	({\mathbbm 1} + \epsilon\Omega) \; \Theta \;
	({\mathbbm 1} - \epsilon\Omega)
\nonumber \\	& = &
	\Theta + \epsilon \left[ \Omega , \Theta \right] + O(\epsilon^2) \; .
\end{eqnarray}		
Now the definition~(\ref{eq:AA10}) implies
\begin{equation}
	\dot\theta = \left( \frac{\rd}{\rd t} R(B) \right) (\Theta) =
	\left( \frac{\rd}{\rd t} R(B) \right) R(B)^{-1} (\theta) \; ,
\end{equation}
so that the above Eq.~(\ref{eq:AA9}) can be expressed in the form
\begin{equation}
	\left( \frac{\rd}{\rd t} R(B) \right) R(B)^{-1} (\theta)
	= r(\Omega)(\theta) \; .
\end{equation}
This is the lifted image of Eq.~(\ref{eq:AA3}), in the same sense as the
Schr\"odinger equation~(\ref{eq:LSG}) in Subsec.~\ref{sec:2b} is the lifted
image of Eq.~(\ref{eq:LIE}). Thus, it would have been possible to deduce
Eq.~(\ref{eq:AA9}) immediately from Eq.~(\ref{eq:AA3}) in a single step.

\section{Damped spin precession}
\label{sec:APB}

\begin{figure}[t]
\centering
\includegraphics[width=1.0\linewidth]{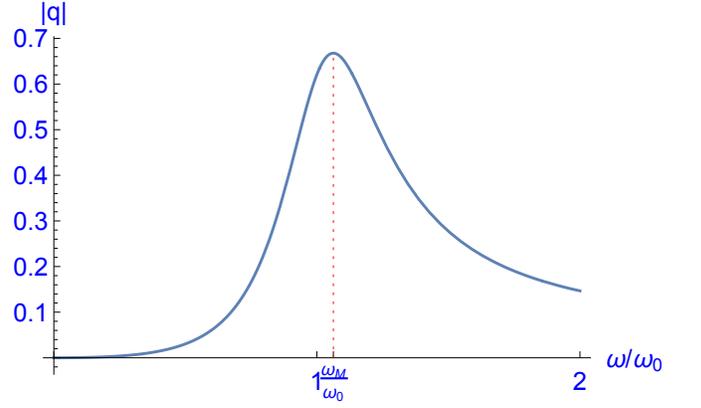}
	\caption{(Color online) Absolute value of the dissipated heat $|q|$ of the classical
	damped, periodically driven spin described by the Landau-Lifshitz
 	equation~(\ref{eq:AB1}) as a function of $\omega/\omega_0$.
	Parameters are $F/\omega_0 = 1/4$, and $g = 0.1$.}
\label{F_12}
\end{figure}

Instead of a driven spin coupled to a heat bath, here we consider a classical
unit spin vector ${\mathbf S}(t)$ under the influence of both a magnetic field
${\mathbf b}(t)$ conforming to Eq.~(\ref{eq:MAF}) and a nonlinear damping
mechanism satisfying an equation of the Landau-Lifshitz
type~\cite{LL35},
\begin{equation}
 	\frac{\rd}{\rd t}{\mathbf S} = -{\mathbf S} \times {\mathbf b} +
	g \, {\mathbf S} \times \left( {\mathbf S} \times {\mathbf b} \right)
 	\; ,
\label{eq:AB1}
\end{equation}
where the parameter $g > 0$ describes the strength of the damping. (Note that
the unusual signs have been chosen here because the spin vector of an electron
points into the direction opposite to that of the magnetic moment.) Analogously
to the elementary case of a classical damped, periodically driven oscillator
it can be shown that asymptotically for $t\to\infty$ the solution of
Eq.~(\ref{eq:AB1}) becomes a rotation about the ${\mathbf b}_z$-axis with the
same frequency as the field ${\mathbf b}(t)$ and a constant phase shift,
\begin{equation}
 	{\mathbf S}(t) = \left( \begin{array}{c}
   	\sqrt{1-z^2} \cos(\omega t - \phi) \\
  	\sqrt{1-z^2} \sin(\omega t - \phi) \\
   	z
 	\end{array} \right) \; .
\label{eq:AB2}
\end{equation}
The functions $z = z(\omega_0,\omega,F,g)$ and
$\phi = \phi(\omega_0,\omega,F,g)$ can be determined analytically, but are
too lengthy to be reproduced here. Let
\begin{equation}
 	E(t) = {\mathbf S}(t) \cdot {\mathbf b}(t)
\end{equation}
be the energy of the spin in the field; this function $E(t)$ will be a periodic
function of~$t$ for the special solution~(\ref{eq:AB2}). Differentiating, we
find
\begin{eqnarray}
  	\dot{E} & = & \dot{\mathbf S}\cdot {\mathbf b }
	+ {\mathbf S}\cdot \dot{\mathbf b }
\nonumber \\	& = &
	-({\mathbf S}\times {\mathbf b}) \cdot {\mathbf b }
        +g \big({\mathbf S} \times ({\mathbf S} \times {\mathbf b}) \big)
	\cdot{\mathbf b } + {\mathbf S} \cdot \dot{\mathbf b }
\nonumber \\	& = &
	g\big( ({\mathbf S}\cdot{\mathbf b})^2 - {\mathbf b}^2 \big)
	+ {\mathbf S} \cdot \dot{\mathbf b } \; .
\label{eq:AB4}
\end{eqnarray}
Integrating $ \dot{E}$ over one period gives~$0$, implying that the time
averages of the final two terms in Eq.~(\ref{eq:AB4}) must cancel. It is
plausible to regard the first term
\begin{equation}
	q \equiv
	g\big( ({\mathbf S} \cdot{\mathbf b})^2 - {\mathbf b}^2 \big)
	\le 0
\label{eq:AB5}
\end{equation}	
as the heat loss of the driven spin due to the damping, and the second term
\begin{equation}
	w \equiv {\mathbf S} \cdot \dot{\mathbf b}
\end{equation}	
as the work performed on the spin by the driving field, so that the time
average of~$w$ must be positive. A short calculation yields
\begin{equation}\label{D5}
  	w = |q| = -F\,\omega\,\sqrt{1-z^2}\,\sin\phi
\end{equation}
which is independent of~$t$, so that time-averaging is not necessary here.
Note that $w \ge 0$ requires that the phase shift~$\phi$ satisfies
$\pi \le \phi \le 2\pi$. In Fig.~\ref{F_12} we have plotted $|q|$ as a
function of $\omega/\omega_0$ for parameters taken from Fig.~\ref{F_8} and
observe qualitative agreement with $r_\infty$ as drawn therein, suggesting
a close connection between the heat-bath model and the phenomenological
Landau-Lifshitz equation~(\ref{eq:AB1}).

A thorough investigation of this connection would require knowledge of the
relation between the dissipation constant~$g$ and the parameters of the
heat-bath model. While a rigorous derivation of that relation is beyond the
scope of this Appendix, one may obtain insight from a simple dimensional
argument: Comparing the factors of dimension 1/time which set the scale for
the relaxation rate implied by Eq.~(\ref{eq:AB1}) on the one hand, and for
the rates (\ref{eq:GAL}) and (\ref{eq:GAH}) on the other, we deduce
\begin{equation}
	g | {\mathbf b} | \; \sim \; \Gamma_0 = 2\pi\gamma^2 J_0 \;  ,
\end{equation}
where the symbol ``$\sim$'' is supposed to indicate ``equality up to
dimensionless numbers of order one''. 	

The rate $|q|$ of dissipated heat can then be expressed in two different
ways. According to Eq.~(\ref{eq:AB5}), one has
\begin{equation}
	| q | \; \sim \; g | {\mathbf b} |^2
	\; =\; g (F^2 + \omega_0^2) \; .
\label{eq:AB9}
\end{equation}	
On the other hand, the maximum dimensionless scaled dissipation rate $r_m$
in the classical limit $s \to \infty$ is given by Eq.~(\ref{eq:IMR}),
implying the estimate
\begin{eqnarray}
	R_m & = & \omega_0 \, \Gamma_0 \, r_m
\nonumber \\	& \sim &
	\Gamma_0 \left( \omega_0 + \sqrt{F^2 + \omega_0^2} \right)
\nonumber \\	& \sim &
	g | {\mathbf b} | \left( \omega_0 + \sqrt{F^2 + \omega_0^2} \right)
\nonumber \\	& \sim &
	g \left( F^2 + \omega_0^2 + \omega_0 \sqrt{F^2 + \omega_0^2} \right)	
\end{eqnarray}	   	
for the full rate. This is compatible with the above relation~(\ref{eq:AB9}),
again hinting at an intrinsic connection between the heat-bath model studied
in the main text, and the Landau-Lifshitz equation~(\ref{eq:AB1}), so that
the remarkable similarity between Figs.~\ref{F_8} and~\ref{F_12} is no
coincidence.


\newpage

\begin{thebibliography}{99}


\bibitem{Zeldovich66}
	Ya. B. Zel'dovich,
	{\em The quasienergy of a quantum-mech\-anical system subjected to a
	periodic action\/},
	J. Exptl. Theoret. Phys. (U.S.S.R.) {\bf 51}, 1492 (1966)
	[Sov. Phys. JETP {\bf 24}, 1006 (1967)].

\bibitem{Sambe73}
	H. Sambe,	
	{\em Steady states and quasienergies of a quantum-mechanical system
	in an oscillating field\/},
	Phys. Rev. A {\bf 7}, 2203 (1973).
	
\bibitem{FainshteinEtAl78}
	A. G. Fainshtein, N. L. Manakov, and L. P. Rapoport,
	{\em Some general properties of quasi-energetic spectra of
	quantum systems in classical monochromatic fields\/},
	J. Phys. B: Atom. Molec. Phys. {\bf 11}, 2561 (1978).


\bibitem{BlumelEtAl91}
	R. Bl\"umel, A. Buchleitner, R. Graham, L. Sirko, U. Smilansky,
	and H. Walther,
	{\em Dynamical localization in the microwave interaction of Rydberg
	atoms: The influence of noise\/},
	Phys. Rev. A {\bf 44}, 4521 (1991).
	
\bibitem{GrifoniHanggi98}
	M. Grifoni and P. H\"anggi,
	{\em Driven quantum tunneling\/},
	Phys. Rep. {\bf 304}, 229 (1998).
	
\bibitem{GasparinettiEtAl13}
	S. Gasparinetti, P. Solinas, S. Pugnetti, R. Fazio, and J. P. Pekola,
	{\em Environment-governed dynamics in driven quantum systems\/},
	Phys. Rev. Lett. {\bf 110}, 150403 (2013).
	
\bibitem{StaceEtAl13}
	T. M. Stace, A. C. Doherty, and D. J. Reilly,
	{\em Dynamical steady states in driven quantum systems\/},
	 Phys. Rev. Lett. {\bf 111}, 180602 (2013).	
		
\bibitem{ZhangEtAl17}
	J. Zhang, P. W. Hess, A. Kyprianidis, P. Becker, A. Lee, J. Smith,
	G. Pagano, I.-D. Potirniche, A. C. Potter, A. Vishwanath, N. Y. Yao,
	and C. Monroe,
	{\em Observation of a discrete time crystal\/},
	Nature {\bf 543}, 217 (2017).
	
\bibitem{ChoiEtAl17}
	S. Choi, J. Choi, R. Landig, G. Kucsko, H. Zhou, J. Isoya, F. Jelezko,
	S. Onoda, H. Sumiya, V. Khemani, C. von Keyserlingk, N. Y. Yao,
	E. Demler, and M. D. Lukin,
	{\em Observation of discrete time-crystalline order in a disordered
	dipolar many-body system\/},
	Nature {\bf 543}, 221 (2017).	
	
	
\bibitem{Kohn01}
	W. Kohn,
	{\em Periodic Thermodynamics\/},
	J. Stat. Phys. {\bf 103}, 417 (2001).
	
\bibitem{BreuerEtAl00}
	H.-P. Breuer, W. Huber, and F. Petruccione,
	{\em Quasistationary distributions of dissipative nonlinear quantum
	oscillators in strong periodic driving fields\/},
	Phys. Rev. E {\bf 61}, 4883 (2000).	

\bibitem{KetzmerickWustmann10}
	R. Ketzmerick and W. Wustmann,
	{\em Statistical mechanics of Floquet systems with regular and chaotic
	states\/},
	Phys. Rev. E {\bf 82}, 021114 (2010).
	
\bibitem{HoneEtAl09}
	D. W. Hone, R. Ketzmerick, and W. Kohn,
	{\em Statistical mechanics of Floquet systems: The pervasive problem
	of near-degeneracies\/},
	Phys. Rev. E {\bf 79}, 051129 (2009).
	
\bibitem{BulnesCuetaraEtAl15}
	G. Bulnes Cuetara, A. Engel, and M. Esposito,
	{\em Stochastic thermodynamics of rapidly driven systems\/},
	New J. Phys. {\bf 17}, 055002 (2015).	

\bibitem{ShiraiEtAl15}
	T. Shirai, T. Mori, and S. Miyashita,
	{\em Condition for emergence of the Floquet-Gibbs state in
	periodically driven open systems\/},
	Phys. Rev. E {\bf 91}, 030101(R) (2015).
	
\bibitem{Liu15}
	D. E. Liu,
	{\em Classification of the Floquet statistical distribution for
	time-periodic open systems\/},
	Phys. Rev. B {\bf 91}, 144301 (2015).
	
\bibitem{IadecolaEtAl15a}
	T. Iadecola, and C. Chamon,
	{\em Floquet systems coupled to particle reservoirs\/},
	Phys. Rev. B {\bf 91}, 184301 (2015).	

\bibitem{IadecolaEtAl15}
	T. Iadecola, T. Neupert, and C. Chamon,
	{\em Occupation of topological Floquet bands in open systems\/},
	Phys. Rev. B {\bf 91}, 235133 (2015).
	
\bibitem{SeetharamEtAl15}
	K. I. Seetharam, C.-E. Bardyn, N. H. Lindner, M. S. Rudner, and
	G. Refael,
	{\em Controlled population of Floquet-Bloch states via coupling
	to Bose and Fermi baths\/},
	Phys. Rev. X {\bf 5}, 041050 (2015).
	
\bibitem{VorbergEtAl15}
	D. Vorberg, W. Wustmann, H. Schomerus, R. Ketzmerick, and A. Eckardt,
	{\em Nonequilibrium steady states of ideal bosonic and fermionic
	quantum gases\/},
	Phys. Rev. E {\bf 92}, 062119 (2015).	
	
\bibitem{VajnaEtAl16}
	S. Vajna, B. Horovitz, B. D\'ora, and G. Zar\'and,
	{\em Floquet topological phases coupled to environments and the
	induced photocurrent\/},
	Phys. Rev. B {\bf 94}, 115145 (2016).	
	
\bibitem{RestrepoEtAl16}
	S. Restrepo, J. Cerrillo, V. M. Bastidas, D. G. Angelakis, and
	T. Brandes,
	{\em Driven open quantum systems and Floquet stroboscopic dynamics\/},
	Phys. Rev. Lett. {\bf 117}, 250401 (2016).
					 		
\bibitem{LazaridesMoessner17}
	A. Lazarides and R. Moessner,
	{\em Fate of a discrete time crystal in an open system\/},
	Phys. Rev. B {\bf 95}, 195135 (2017).
			
\bibitem{SeetharamEtAl19}
	K. I. Seetharam, C.-E. Bardyn, N. H. Lindner, M. S. Rudner,
	and G. Refael,
	{\em Steady states of interacting Floquet insulators\/},
	Phys. Rev. B {\bf 99}, 014307 (2019).
						
		
\bibitem{Rabi37}
	I. I. Rabi,
	{\em Spin quantization in a gyrating magnetic field\/},
	Phys. Rev. {\bf 51}, 652 (1937).

\bibitem{Brillouin27}
	L. Brillouin,
	{\em Les moments de rotation et le magn\'etisme dans la
	m\'ecanique ondulatoire\/},
	J. Phys. Radium  {\bf 8}, 74 (1927).
	
\bibitem{Henry52}
	W. E. Henry,
	{\em Spin Paramagnetism of $\mbox{Cr}^{+++}$, $\mbox{Fe}^{+++}$, and
	$\mbox{Gd}^{+++}$ at Liquid Helium Temperatures and in Strong Magnetic
	Fields\/},
	Phys. Rev. {\bf 88}, 559 (1952).
	
\bibitem{FowlerGuggenheim39}
	R. H. Fowler and E. A. Guggenheim,
	{\em Statistical Thermodynamics\/}
	(Cambridge University Press, Cambridge, 1939).	
	
\bibitem{Pathria11}
	For a modern exposition see, e.g.,
	R. K. Pathria,
	{\em Statistical Mechanics\/}
	(Academic Press, New York; 3rd edition, 2011).
	
\bibitem{LangemeyerHolthaus14}
	M. Langemeyer and M. Holthaus,
	{\em Energy flow in periodic thermodynamics\/},
	Phys. Rev. E {\bf 89}, 012101 (2014).
	
\bibitem{Schmidt18}
	H.-J. Schmidt,
	{\em The Floquet theory of the two-level system revisited\/},
	Z. Naturforsch. A {\bf 73}, 705 (2018).
	
\bibitem{LaLiQM81}
	L. D. Landau and E. M. Lifshitz,
	{\em Quantum Mechanics: Non-Relativistic Theory\/}, \S~57
	(Butterworth-Heinemann, 3rd revised edition, reprinted 1981).		
	
\bibitem{Hioe87}
	F. T. Hioe,
	{\em $N$-level quantum systems with $SU(2)$ dynamic symmetry\/},
	J. Opt. Soc. Am. B {\bf 4}, 1327 (1987).

\bibitem{PokrovskySinitsyn04}
	V. L. Pokrovsky and N. A. Sinitsyn,
	{\em Spin transitions in time-dependent regular and random magnetic
	fields\/},	
	Phys. Rev. B {\bf 69}, 104414 (2004).

\bibitem{Hall15}
	See, e.g.,
	B. C. Hall,
	{\em Lie Groups, Lie Algebras, and Representations: An Elementary
	Introduction\/}, Graduate Texts in Mathematics {\bf 222}
	(Springer, New York; 2nd edition, 2015).

\bibitem{HolthausJust94}		
	M. Holthaus and B. Just,
        {\em Generalized $\pi$-pulses\/},
	Phys. Rev. A {\bf 49}, 1950 (1994).		

\bibitem{Shirley65}
	J. H. Shirley,
	{\em Solution of the Schr\"odinger equation with a Hamiltonian
	periodic in time\/},
	Phys. Rev. {\bf 138}, B~979 (1965).

\bibitem{Salzman74}
	W. R. Salzman,
	{\em Quantum mechanics of systems periodic in time\/},
	Phys. Rev. A {\bf 10}, 461 (1974).
	
\bibitem{GesztesyMitter81}
	F. Gesztesy and H. Mitter,
	{\em A note on quasi-periodic states\/},
	J. Phys. A: Math. Gen. {\bf 14}, L79 (1981).
	
\bibitem{Holthaus16}		
	M. Holthaus,
	{\em Floquet engineering with quasienergy bands of periodically driven
	optical lattices\/},
	J. Phys. B: At. Mol. Opt. Phys. {\bf 49}, 013001 (2016).
		
\bibitem{ExplainV}
	Note a minor deviation from Ref.~\cite{LangemeyerHolthaus14},
	where $V= \gamma\,\sigma_x = 2\gamma\,s_x$.
	
\bibitem{DiermannEtAl19}
	O. R. Diermann, H. Frerichs, and M. Holthaus,
	{\em Periodic thermodynamics of the parametrically driven harmonic
	oscillator\/},
	Phys. Rev. E {\bf 100}, 012102 (2019). 	
	
\bibitem{EatonEtAl10}
	G. R. Eaton, S. S. Eaton, D. P. Barr, and R. T. Weber,
	{\em Quantitative EPR\/}
	(Springer-Verlag, Wien; 2010).
	
\bibitem{GasparinettiEtAl14}
	S. Gasparinetti, P. Solinas, A. Braggio, and M. Sassetti,
	{\em Heat-exchange statistics in driven open quantum systems\/},
	New J. Phys. {\bf 16}, 115001 (2014).
	
\bibitem{LL35}		
	L.D. Landau and E.M. Lifshitz,
	{\em   Theory of the dispersion of magnetic permeability in ferromagnetic bodies\/},
    Phys. Z. Sowjetunion  {\bf 8}, 153 (1935).


\bibitem{PetukhovEtAl05}		
	K. Petukhov, W. Wernsdorfer, A.-L. Barra, and V. Mos\-ser,
	{\em Resonant photon absorption in Fe$_8$ single-molecule magnets
	detected via magnetization measurements\/},
	Phys. Rev. B {\bf 72}, 052401 (2005).
			
\end{thebibliography}
\end{document}